\documentclass[prd,amsmath,amssymb,twocolumn]{revtex4-1}

\usepackage{graphicx}
\usepackage{color}
\usepackage{amsmath}
\usepackage{amssymb}

\usepackage{bm}
\usepackage{bbm}
\usepackage{cancel}
\usepackage{hyperref}

\newcommand{\CO}{\mathbb{C}\otimes\mathbb{O}}
\newcommand{\CHO}{\mathbb{C}\otimes\mathbb{H}\otimes\mathbb{O}}
\newcommand{\RCHO}{\mathbb{R}\otimes\mathbb{C}\otimes\mathbb{H}\otimes\mathbb{O}}
\newcommand{\A}{\mathbb{A}}

\newcommand{\CLeight}{\mathbb{C}l(8)}

\newcommand{\CLtwo}{\mathbb{C}l(2)}

\newcommand{\CLten}{\mathbb{C}l(10)}

\newcommand{\gsm}{\mathfrak{su}(3)_C \oplus \mathfrak{su}(2)_L \oplus \mathfrak{u}(1)_Y}
\newcommand{\gle}{\mathfrak{su}(3)_C  \oplus \mathfrak{u}(1)_Q}

\newcommand{\C}{\mathbb{C}}
\newcommand{\R}{\mathbb{R}}

\newcommand{\OHCR}{\mathbb{O}\oplus \mathbb{H} \oplus \mathbb{C}\oplus \mathbb{R}}

\usepackage{xcolor}
\definecolor{navyblue}{RGB}{10,23,173}
\definecolor{navygreen}{RGB}{50,125,82}

\usepackage{tikz}

\newcommand{\grayunderlinedmath}[1]{%
  \tikz[baseline=(X.base)]{
    \node[inner sep=0pt, outer sep=0pt] (X) {$#1$};
    \draw[gray!35, line width=0.35pt]
      ([yshift=-0.45ex]X.south west) -- ([yshift=-0.45ex]X.south east);
  }%
}

\begin{document}

\title{Standard Model Symmetries\vspace{2mm} \\and the \vspace{2mm}\\Nested Embeddings of $\R\subset\C\subset\mathbb{H}\subset \mathbb{O}$}

\author{ N. Furey}
\affiliation{$ $\\  Iris Adlershof, Humboldt-Universit\"{a}t zu Berlin,\\ Zum Grossen Windkanal 2,  Berlin, Germany  \\and\\ Nicolaus Copernicus University, Grudziadzka 5, Toru\'{n}, Poland\vspace{1mm}\\ furey@physik.hu-berlin.de\\HU-EP-26/24}

\begin{abstract}  Where does the Standard Model's internal structure come from?  Treating $\OHCR$ as a module for its own multiplication algebra enables a particular origin story for the Standard Model's pre-Higgs, $\mathfrak{g}_{\textup{SM}}:=\gsm,$ and low energy post-Higgs, $\mathfrak{g}_{\textup{LE}}:=\gle,$  symmetries.   In this model, weak hypercharge and electric charge operators, $Y$ and $Q,$ take on a remarkably simple form:  $\sum \frac{1}{n}\mathbb{I}_{n\times n}$.  

Upon the introduction of Cayley-Dickson imaginary units, this 15$\hspace{.5mm} \R$ dimensional $\OHCR$ embeds naturally as a vector space into several well-studied 16$\hspace{.5mm} \R$ dimensional algebras, which we generically refer to as $\mathbb{V}.$  With this embedding, the Standard Model's internal symmetries may then be seen to arise in part from the sequence of nested inclusions:  $\R\subset\C\subset\mathbb{H}\subset \mathbb{O}\subset\mathbb{V}.$  In the sedenionic case of $\mathbb{V} = \mathbb{S},$ the full sequence becomes a Cayley-Dickson tower.  We define the notion of \emph{endomorphic models} of particle physics, and connect $\textup{End}_{\R}(\mathbb{V})\simeq Cl(0,8)$ to the earlier ideas of \emph{Bott Periodic Particle Physics.}  We comment on a possible connection between the existence of multiple complex structures and the baryon asymmetry problem.  In closing, we identify an appearance in this model of the fully connected tree of division algebraic Hopf fibrations that starts at $S^{15},$ and \emph{simultaneously} involves the four parallelizable spheres $S^7, S^3,S^1, S^0.$  
 \end{abstract}

\maketitle

\section{Nature's feat of efficiency}

It is no small feat that by using only four bases:  A, C, G, T,  Nature lays the foundation for the entire human genome.  But how exactly can it accomplish this?  Nature accomplishes this not by employing those four bases alone, but rather, by building \emph{sequences} from the bases.

With this efficient encoding in mind, we then define \emph{endomorphic models} of particle physics.  An endomorphic model of particle physics is an algebraic model whereby the particle content is not described directly by the algebra itself, but instead, by the algebra's vector-space endomorphisms.  In most cases we will be interested in, these endomorphisms will be generated by \emph{sequences} of algebraic elements multiplying themselves \citep{Gen}-\citep{Z5}.  Our current aim is to yield the Standard Model's full particle content, including gauge bosons, Higgs, and three generations of fermions, using  the multiplicative sequences of only some low-dimensional algebra.

But which low-dimensional algebra are we then to choose?  In a significant change of course, we direct our attention away from our previous algebras of study, and toward the unsung algebra $\OHCR$.  That is, the direct sum of the octonions, $\mathbb{O},$ the quaternions, $\mathbb{H},$ the complex numbers, $\C$ and the real numbers, $\R.$ $\OHCR$ itself is 15$\hspace{.5mm}\R$ dimensional (and is to be defined precisely in Section~\ref{conv}).  

As we will see, upon the inclusion of Cayley-Dickson imaginary units, $\OHCR$ may be embedded as a vector space into several commonly studied 16$\hspace{.5mm}\R$ dimensional algebras. We label these 16$\hspace{.5mm}\R$ dimensional algebras collectively as $\mathbb{V}$.  Towards the end of this article, we show how the embedding of $\OHCR$ into $\mathbb{V}$ correlates with a particular sequence of nested Cayley-Dickson embeddings $\R\subset\C\subset\mathbb{H}\subset\mathbb{O}\subset\mathbb{V}.$  

The endomorphism algebra $\textup{End}_{\R}(\mathbb{V})$ is of special interest for more than one reason.  At 256$\hspace{.5mm}\R$ degrees of freedom, it is approximately the same size as the Standard Model's common 244 off-shell real-component counts.  This counting includes gauge bosons, Higgs, and three generations of fermions - augmented to include three right-handed neutrinos.  (For this article, it should be noted that we use a complex structure within the Clifford algebra $Cl(0,8),$ thereby reducing it to $M_8(\C):$ a starting point that allows for a chiral model with on-shell counting.) 

Furthermore, $\textup{End}_{\R}(\mathbb{V})$ is isomorphic to $Cl(0,8)$ as an associative algebra.  The model then gains access to concepts of Bott Periodicity, and the stated proposal of a \emph{Bott Periodic Fock Space}.  See~\citep{Bott1}, \citep{Bott2}, \citep{Z5}, and the last section of this article.  

A Bott Periodic Fock space may be understood in the following way.  Let us identify \emph{single} particle states as being embedded in $Cl(0,8)\simeq M_{16}(\R)$.  Bott Periodicity  tells us that any non-degenerate real Clifford algebra $Cl(s,t)$ may be rewritten as $Cl(s,t)\simeq Cl(s_0,t_0)\otimes M_{16}(\R)^{\otimes n} $ for some $s_0, t_0\in\{0,1,\dots 7\},$ and for some $n\in\mathbb{Z}\geq 0$.  We then identify \emph{multiparticle} states as being embedded in the $M_{16}(\R)^{\otimes n}$ portion of $Cl(s,t),$ where $n,$ the number of factors of $M_{16}(\R),$  corresponds to the number of particles in the multiparticle states.  The direct sum of all Clifford algebras of the same type $(s_0, t_0)$ then may collectively form a \emph{Bott Periodic Fock space,}  $\mathcal{F}_{s_0,t_0}.$  That is, 
\begin{equation}\begin{array}{rll}
\mathcal{F}_{s_0,t_0}:=&&\vspace{2mm}\\
&Cl(s_0,t_0)&\color{gray} |\hspace{.5mm} 0 \hspace{.5mm}\rangle \vspace{2mm}\\
\oplus &Cl(s_0,t_0)\otimes M_{16}(\R)&\color{gray} | \hspace{.5mm}p_{i_1} \hspace{.5mm}\rangle \vspace{2mm}\\
\oplus &Cl(s_0,t_0)\otimes M_{16}(\R)\otimes M_{16}(\R)  \hspace{1cm}&\color{gray} | \hspace{.5mm}p_{j_1} \hspace{.5mm}\rangle |\hspace{.5mm} p_{j_2} \hspace{.5mm}\rangle\vspace{2mm}\\
&\hspace{.7cm}\vdots&\hspace{2mm}\color{gray}\vdots\color{black}
\end{array}\end{equation}
\noindent For completeness, if desired, one may define $\mathcal{F}$ as the finite direct sum over all possible Clifford algebra types $(s_0, t_0)$ so as to include the full set of all non-degenerate real Clifford algebras, $\{Cl(s,t)\}:$  
\begin{equation} \mathcal{F}:=\bigoplus_{s_0, t_0=0}^7 \mathcal{F}_{s_0,t_0}.
\end{equation}
Of course, further details such as (anti-)symmetrization of tensor powers, etc.,  will need to be considered in order to connect to standard QFT.

It is clear that $Cl(0,8)\simeq M_{16}(\R)$ on its own does not appear to have the substructure necessary to explain why it should splinter into the many particle irreps we see in the Standard Model.  In a recent seminar \citep{Bott2}, it was shown how much of this substructure could result from multiplication by certain octonionic imaginary units, acting as Clifford volume elements.  However, it had been known for quite some time, e.g. \citep{mg16}, \citep{Bott2}, \citep{Z5}, that strategically placed quaternionic structure was also needed.

\begin{figure}[h]
\begin{center}
\includegraphics[width=8cm]{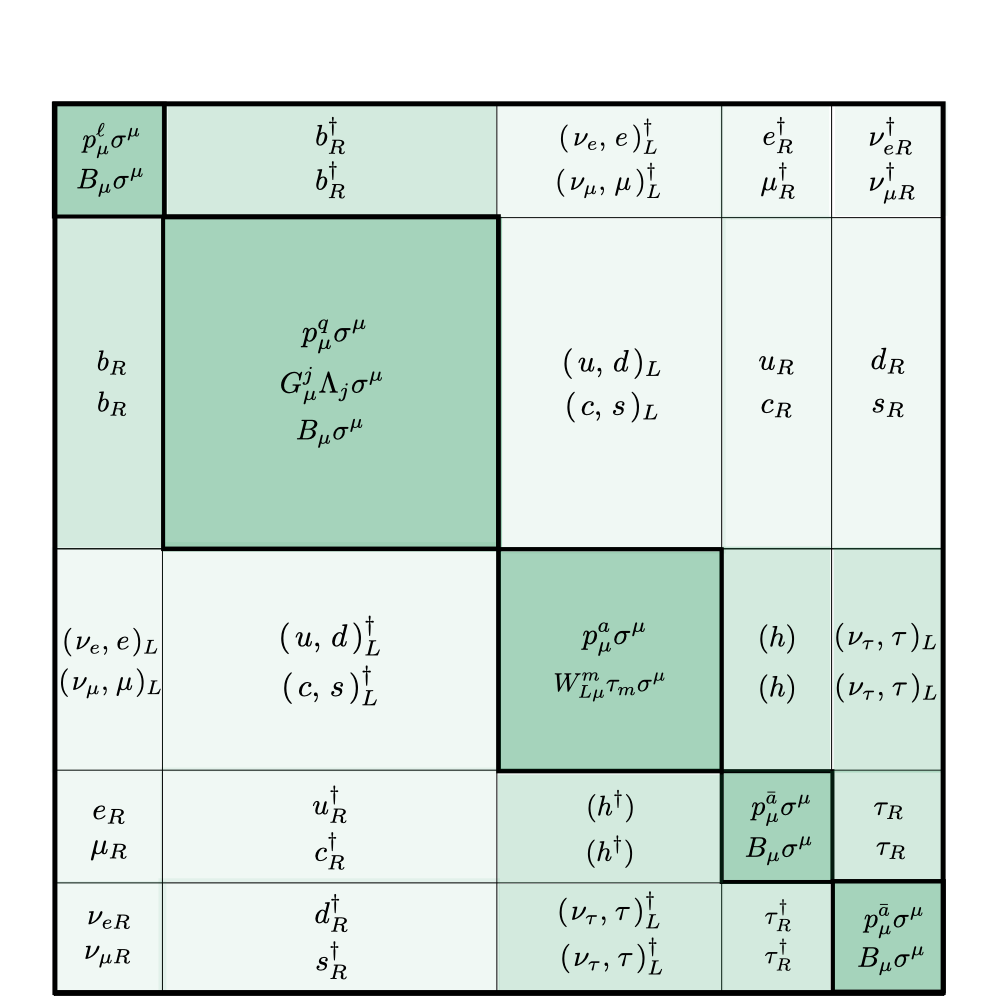}
\caption{\label{map}  Decomposition of  the $16\times 16$ hermitian matrices, $\mathcal{H}_{16}(\C)\subset\CLeight,$ into particle representations according to~\citep{Z5}.  These particle representations mirror the internal behaviour of the Standard Model's gauge bosons, Higgs, and three generations of quarks and leptons -- with exception to those representations involving the top quark.  With this said, it should be noted that the model's extra $b_R',$ $h'$ and $\tau_R'$ internal representations may be combined so as to recover these missing top quark irreps.   It is then natural to question whether the operators describing the heaviest quarks of the Standard Model may in fact be composite.}
\end{center}\end{figure}

This article is the result of an effort to understand why $Cl(0,8)$ might fragment into particle irreps familiar from the Standard Model.  It is part of a long-running research project with a history detailed in~\citep{Z5}.  For the reader's convenience, we reproduce its main diagram here in Figure~(\ref{map}).  With this said, the results found in this current manuscript are general, and may apply to many division algebraic models of particle physics.  The article~\citep{Z5} is not a prerequisite to understanding this text.

\section{Connected works}

That the real numbers, complex numbers, and quaternions are central to modern physics has been known since the dawn of modern physics itself.  However, Nature's relationship with the octonions took longer to recognize.  In the early 1970s, an $\mathfrak{su}(3)$ action was applied to the split octonions by G\"{u}naydin and G\"{u}rsey, \citep{GGquarks}, thereby identifying quark structure.  

Since then, the literature on Cayley-Dickson algebra models of particle physics has broadened substantially.  This includes, but is not limited to models of pure octonions, e.g. Silagadze \citep{Silagadze}, Bryant \citep{bryant}, Gording and Schmidt-May \citep{gsm}, Lasenby \citep{Lasenby}, related split algebras, e.g. Singh and Vaibhav \citep{singh_v}, Penrose \citep{Penrose},  exceptional Lie algebras,  e.g. Chester, Rios, Marrani, \citep{Che2020}, Manogue, Dray, Wilson \citep{Man2022}, Boyle \citep{boyle1}, the exceptional Jordan algebra, e.g. Silagadze \citep{Silagadze}, Manogue and Dray \citep{Man2022}, Dubois-Violette and Todorov \citep{DVtod}, Chester, Marrani, Corradetti, Aschheim, Irwin, \citep{Che2023},  Baez and Schwahn \citep{BS},    the 16-$\R$ dimensional algebra of the sedenions, $\mathbb{S}$, e.g. L\~{o}hmus, Paal, Sorgsepp \citep{NAAP}, K\"{o}plinger \citep{Jens_sed}, Castro Perelman \citep{Carlos2019sed}, Masi \citep{masi}, and complex sedenions, e.g. Weng, \citep{Weng}, Gillard and Gresnigt \citep{niels_sedenion}, $\RCHO,$ e.g. Dixon \citep{Dixon_recent},  Furey and Hughes \citep{dasb},  related braid models, e.g. Chester, Arsiwalla, Kauffman, \citep{lou}, and models of non-commutative geometry, e.g. Connes and Lott \citep{Connes}, Barrett \citep{ncgcliff}.

Since the 1970s, there have been numerous Clifford algebraic models of particle physics constructed e.g. Barducci, Buccella, Casalbuoni,  Lusanna, Sorace \citep{italians}, Trayling and Baylis, \citep{Greg}, Pav\v{s}i\v{c}, \citep{pav}.  Due to Bott periodicity, and size considerations, we will be interested especially in endomorphic models based on the \emph{real} Clifford algebra $Cl(0,8)$.  In 2014, it was proposed by this author to embed Standard Model's bosons and fermions into $Cl(0,8)$'s complexification, $\CLeight\simeq L_{\CHO},$ \citep{Gen}, an idea subsequently developed in \citep{321}, \citep{dasb}, \citep{mg16}, \citep{Z5}.  Gillard and Gresnigt later joined in in 2019 with a related endomorphic $\CLeight$ model, \citep{niels_sedenion}, focussing exclusively on the Standard Model's fermions.  With this said, $\CLeight$ does not underlie real Bott Periodicity, and does not have the correct counting to match that of the Standard Model.  Therefore, the original endomorphic $\CLeight$ proposal in \citep{Gen} was later refined by this author to $Cl(0,8)$ in 2021 \citep{Bott1}, and developed in \citep{dasb}, \citep{Bott2}, \citep{Z5}.  For a $Cl(2,6)\simeq M_8(\mathbb{H})$ model describing one generation of fermions, see~\citep{pav}.

The present article provides a number of new findings.  It introduces a symmetry construction rooted in the largely unknown algebra $\OHCR$.  As we will see, using a new $ \textup{Tr} \hspace{.5mm}\!\bigl(J_{\mathbb{O}}\hspace{.5mm}\ell_{\mathbb O}\bigr)
=
\textup{Tr}\hspace{.5mm} \!\bigl(J_{\mathbb{H}}\hspace{.5mm}\ell_{\mathbb H}\bigr)
=
\textup{Tr}\hspace{.5mm} \!\bigl(J_{\mathbb{C}}\hspace{.5mm}\ell_{\mathbb C}\bigr)$ constraint, it introduces concise expressions for weak hypercharge
\begin{equation} \color{navygreen}Y = \frac{1}{3} P_{\mathbb{O}_2} + \frac{1}{2}P_{\mathbb{H}} + P_{\C},\color{black}
 \end{equation}
\noindent and electric charge
\begin{equation}  \color{navygreen}Q = \frac{1}{3} P_{\mathbb{O}_2} + P_{\mathbb{H}_2} + P_{\C}.\color{black}
 \end{equation}
\noindent We demonstrate how special properties unique to the quaternionic multiplication algebras enable an $\mathfrak{su}(2)_{\textup{L}}\oplus\mathfrak{u}(1)_{\textup{Y}}\mapsto \mathfrak{u}(1)_{\textup{Q}}$ transition familiar from electroweak symmetry breaking.

It is then shown that this $\OHCR$ structure can be embedded as a vector space into several 16$\hspace{.5mm}\R$ dimensional algebras, generically referred to as $\mathbb{V}.$  From here, it is established that the Standard Model's internal symmetries result in part from the nested Cayley-Dickson embeddings $\R\subset\C\subset\mathbb{H}\subset \mathbb{O}\subset\mathbb{V}.$  

We close by demonstrating a connection between this $\R\subset\C\subset\mathbb{H}\subset \mathbb{O}\subset\mathbb{V}$ model and the \emph{entire} tree of division algebraic Hopf fibrations 
\begin{equation}\begin{array}{ccccccccc}
S^{0} &\hookrightarrow& S^1 &\hookrightarrow& S^3 &\hookrightarrow& S^7 &\hookrightarrow& S^{15}\vspace{2mm}\\
&& \downarrow &&\downarrow &&\downarrow &&\downarrow \vspace{2mm}\\
&& S^1 &&S^2 &&S^4 &&S^8 
\end{array}\end{equation}

Due to the generality of endomorphisms, these results may find wide application for many division algebraic models.  In Section~\ref{e8}, we offer one application in the context of $\mathfrak{e}_8.$

\section{Multiplication algebras}

It is often not fully appreciated that multiplying any two algebraic elements together automatically invokes that algebra's \emph{multiplication algebra}.  As an example demonstrated below in equation~(\ref{su3}), the $\mathfrak{su}(3)\subset \mathfrak{der}(\mathbb{O})$ generators used to describe gluons in octonionic theories are not themselves elements of the octonions.  Rather, they are given by \emph{sequences} of octonions being left-multiplied onto octonions.  Stated precisely, these $\mathfrak{su}(3)$ elements then do not reside in $\mathbb{O}$ itself, but instead in  $\mathbb{O}$'s multiplication algebra.  (They close, of course, under the commutator.)  In the case of the octonions, one finds that its left- and right-multiplication algebras each coincide with $\textup{End}_{\R}(\mathbb{O}).$

In octonionic theories, the realization that  gluons should embed in multiplication algebras is one motivation for exploring the possibility that in fact all local bosonic and fermionic particle representations could embed in that same space, \citep{Gen}-\citep{Z5}.  The idea characterizes a class of theories where the degrees of freedom are not described directly by the base algebra \emph{per se}, but rather, by that algebra's endomorphisms.  See endnote~\citep{endom}.

The action of an algebra's endomorphisms on its base algebra is in many ways reminiscent of operators acting on a Hilbert space in quantum field theory.  This suggests that endomorphic models might then naturally capture certain desirable features of QFT.


\subsection{Definitions and example}

Let $\A$ be a (possibly nonassociative) algebra over a field $\mathbb{F}$.  Given $x \in \A$, the symbol $L_x$ denotes the $\mathbb{F}$-linear map $\A \to \A$ given by 
\begin{equation}L_x(y) := xy \hspace{5mm} \forall y \in \A.
\end{equation} 
\noindent Similarly, the symbol $R_x$ denotes the $\mathbb{F}$-linear map $\A \to \A$ given by 
\begin{equation}R_x(y) := yx  \hspace{5mm} \forall y \in \A.
\end{equation} 

We then define $\A$'s \emph{left multiplication algebra}, $L_{\A},$ to be the subalgebra of $\textup{End}_{\mathbb{F}}(\A)$ generated by $\{L_x \mid x \in \A\}$.  We  define $\A$'s \emph{right multiplication algebra}, $R_{\A},$ as the subalgebra of $\textup{End}_{\mathbb{F}}(\A)$ generated by $\{R_x \mid x \in \A\}$.  Finally, we define $\A$'s \emph{full multiplication algebra}, $B_{\A},$ to be the subalgebra of $\textup{End}_{\mathbb{F}}(\A)$ generated by both $\{L_x \mid x \in \A\}$ and $\{R_{x'} \mid x' \in \A\}.$   It should be clear that the dimension of $\A$ need not match the dimension of any of its multiplication algebras.

Composition of maps naturally serves as multiplication within the algebras $L_{\A},$ $R_{\A},$ and $B_{\A}$.  Since the composition of maps is associative, $L_{\A},$ $R_{\A},$ and $B_{\A}$ each necessarily constitutes an associative algebra, even when $\A$ is not itself associative.  It can be seen that $L_a\circ L_b(y) = a(by)$, and $R_a\circ R_b(y) = (yb)a$ for all $a,b,y\in\A.$  However, for non-associative algebras $\A,$ it cannot be assumed that $L_a\circ L_b(y) = L_{ab}(y),$ or that $R_a\circ R_b(y) = R_{ba}(y).$

We are now in a position to present G\"{u}naydin and G\"{u}rsey's example, \citep{GGquarks}, of $\mathfrak{su}(3)$ symmetries as elements within the octonionic left multiplication algebra, $L_{\mathbb{O}}$:
\begin{equation}\begin{array}{lll} \label{su3}
\delta_{\mathfrak{su}(3)}&=&  \frac{\rho_1}{4}\left(L_{34}-L_{15} \right) + \frac{\rho_2}{4}\left(L_{14}+L_{35} \right) \vspace{2mm} \\
 &+& \frac{\rho_3}{4}\left(L_{13}-L_{45} \right) -\frac{\rho_4}{4}\left(L_{25}+L_{46} \right) \vspace{2mm} \\
 &+& \frac{\rho_5}{4}\left(L_{24}-L_{56} \right) -\frac{\rho_6}{4}\left(L_{16}+L_{23} \right)\vspace{2mm} \\
 &-&  \frac{\rho_7}{4}\left(L_{12}+L_{36} \right) - \frac{\rho_8}{4\sqrt{3}}\left(L_{13}+L_{45}  -  2L_{26} \right),
\end{array}\end{equation}
\noindent where $\rho_k\in\R,$ and $L_{ij}$ is shorthand for $L_i\circ L_j = L_{e_i}\circ L_{e_j}.$  Here $e_i$ and $e_j$ are octonionic imaginary units with $i,j\in\{1, \dots 7\}.$

\subsection{$\R,$ $\C,$ $\mathbb{H},$ and $\mathbb{O}$ conventions\label{conv}}

In an effort to maintain notational consistency with previous work, we write a generic complex number as $r_0 +r_1i$, where $r_0,r_1\in\R,$ and $i^2 = -1$. 

We write a generic quaternion as $r_0 +r_m \epsilon_m$, for $m\in\{1,2,3\},$ where $r_0,r_m\in\R.$ Quaternionic imaginary units multiply as $\epsilon_m\epsilon_n = -\delta_{mn} + \varepsilon_{mnp}\epsilon_p$ where $\varepsilon_{mnp}$ is the usual totally anti-symmetric tensor with $\varepsilon_{123} = 1$.  

We write a generic octonion as $r_0 +r_j e_j$, for $j\in\{1,2,\dots 7\},$ where $r_0,r_j\in\R.$ Octonionic imaginary units multiply as $e_ie_j =-\delta_{ij}+ f_{ijk} e_k$, where $f_{ijk}$ is again a totally antisymmetric tensor with $f_{ijk}=1$ when $ijk\in \{124, 235, 346, 457, 561, 672, 713\}$.  Those remaining values of $f_{ijk}$ not determined by anti-symmetry are otherwise zero.  

Throughout this article, tensor products will be understood to be over $\R$, unless otherwise stated. 

Finally, we define $\OHCR$ as a 15$\hspace{.5mm}\R$ dimensional vector space, together with componentwise multiplication.  An element of $\OHCR$ is written $a:=(o, h, c, r)$ with $o\in\mathbb{O},$ $h\in\mathbb{H},$ $c\in\mathbb{C},$ $r\in\mathbb{R}.$  Then multiplication between any $a_1$ and $a_2$ $\in \OHCR$  is carried out as 
\begin{equation} (o_1, h_1, c_1, r_1) \cdot (o_2, h_2, c_2, r_2) = (o_1o_2, \hspace{.5mm}h_1h_2, \hspace{.5mm}c_1c_2, \hspace{.5mm}r_1r_2).
\end{equation}

\subsection{$\R,$ $\C,$ $\mathbb{H},$ and $\mathbb{O}$ multiplication algebras}

The various multiplication algebras of $\R,$ $\C,$ $\mathbb{H},$ and $\mathbb{O}$ exhibit relations that can at times be surprising.  These features will ultimately allow us to arrive at the Standard Model's internal symmetries in the upcoming sections.  Furthermore, features unique to quaternionic multiplication algebras will be seen to enable an $\mathfrak{su}(2)_{\textup{L}}\oplus\mathfrak{u}(1)_{\textup{Y}}\mapsto \mathfrak{u}(1)_{\textup{Q}}$ transition familiar from electroweak symmetry breaking.

Given the set up of subsection~\ref{conv}, readers may confirm the following \emph{elementwise} properties 
\begin{equation}\begin{array}{lll}
\R: &\hspace{5mm} L_a = R_a &\hspace{5mm} \forall a \in  \R, \vspace{2mm}\\
\C: &\hspace{5mm} L_a = R_a & \hspace{5mm}\forall a \in  \C, \vspace{2mm}\\
\mathbb{H}: & \hspace{5mm}L_a \neq R_a & \hspace{5mm}\textup{unless } a \in  \R\subset\mathbb{H}, \vspace{2mm}\\
\mathbb{O}: & \hspace{5mm}L_a \neq R_a &\hspace{5mm} \textup{unless } a \in  \R\subset\mathbb{O}.
\end{array}\end{equation}

\emph{As operator algebras}, on the other hand, readers may verify that
\begin{equation}\begin{array}{ll}
\R: &\hspace{5mm} L_{\R} \simeq  R_{\R}, \vspace{2mm}\\
\C: &\hspace{5mm} L_{\C} \simeq  R_{\C}, \vspace{2mm}\\
\mathbb{H}: & \hspace{5mm}L_{\mathbb{H}} \cap R_{\mathbb{H}} \simeq  L_{\R\hspace{.2mm}\subset\hspace{.2mm}\mathbb{H}}\simeq  R_{\R\hspace{.2mm}\subset\hspace{.2mm}\mathbb{H}}, \vspace{2mm}\\
\mathbb{O}: & \hspace{5mm}L_{\mathbb{O}} \simeq  R_{\mathbb{O}}.
\end{array}\end{equation}

In particular, we point out that $L_{\mathbb{H}}$ and $R_{\mathbb{H}}$ supply two (mostly) distinct sets of endomorphisms on $\mathbb{H}$.  This may come as no surprise, given that $\mathbb{H}$ is non-commutative.  However, $\mathbb{O}$ is also non-commutative, and yet one finds that $L_{\mathbb{O}} \simeq  R_{\mathbb{O}}.$

This puzzle may be resolved by noting that for right multiplication of octonionic imaginary units, $e_i,$ with $i\in\{1\dots 7\},$
\begin{equation} R_i = \frac{1}{2}\left( -L_i + L_{i+1\hspace{.7mm}i+3}+ L_{i+2\hspace{.7mm}i+6}+ L_{i+4\hspace{.7mm}i+5}  \right),
\end{equation}
\noindent where indices are evaluated mod 7.  That is, every octonionic right-multiplication map may be expressed as some element of $L_{\mathbb{O}}$, and, by symmetry, every octonionic left-multiplication map may be expressed as some element of $R_{\mathbb{O}}$.

The full multiplication algebras, $B_{\A},$ relate to the others as
\begin{equation}\begin{array}{ll}\label{full}
\R: &\hspace{5mm} B_{\R}\simeq L_{\R} \simeq  R_{\R}, \vspace{2mm}\\
\C: &\hspace{5mm} B_{\C}\simeq L_{\C} \simeq  R_{\C}, \vspace{2mm}\\
\mathbb{H}: & \hspace{5mm}B_{\mathbb{H}}  \simeq  L_{\mathbb{H}}\otimes_{\R}R_{\mathbb{H}}, \vspace{2mm}\\
\mathbb{O}: & \hspace{5mm}B_{\mathbb{O}} \simeq  L_{\mathbb{O}} \simeq  R_{\mathbb{O}}.
\end{array}\end{equation}
\noindent Readers should note the unique properties of the quaternionic multiplication algebras relative to the others.

\section{Clifford algebras and \\imaginary volume elements\label{imagine}}

For each case of $L_{\A},$ $R_{\A},$ and $B_{\A},$ it is possible to identify certain Clifford algebras to which they are isomorphic (as associative algebras).  We will use the convention that $Cl(s,t)$ has $s$ generators that square to 1, and $t$ generators that square to -1. 

For reasons soon to become clear, we will be interested only in Clifford algebras whose volume element squares to -1.  That is, whenever they indeed have a nontrivial volume element.  Such a set up has long been central in the development of geometric algebra, e.g. \citep{Lasenby}.  See also~\citep{dasb}, \citep{roadI}, \citep{roadIII}.    Explicitly,
\begin{equation}\begin{array}{ll}\label{endcliffs}
\textup{End}_{\R}(\R)\simeq &L_{\R}\simeq R_{\R}\simeq B_{\R}\simeq \R \color{gray}\simeq Cl(0,0),\vspace{2mm}\\
\textup{End}_{\C}(\C)\simeq &L_{\C}\simeq R_{\C}\simeq B_{\C}\simeq \C \color{gray}\simeq \C l(0),\vspace{2mm}\\
\textup{End}_{\R}(\mathbb{H})\simeq &B_{\mathbb{H}}\simeq M_4(\R)\color{navyblue}\simeq Cl(3,1),\vspace{2mm}\\
&L_{\mathbb{H}}\simeq \mathbb{H}\color{navyblue}\simeq Cl(0,2),\vspace{2mm}\\
&R_{\mathbb{H}}\simeq \mathbb{H}^{op}\simeq \mathbb{H}\color{navyblue}\simeq Cl(0,2),\vspace{2mm}\\
\textup{End}_{\R}(\mathbb{O})\simeq &L_{\mathbb{O}}\simeq R_{\mathbb{O}}\simeq B_{\mathbb{O}}\simeq M_8(\R)\color{navyblue}\simeq Cl(0,6).\color{black}
\end{array}\end{equation}
\noindent In order to treat all algebras as real algebras, one may take the multiplication algebras of $\C$ to be $\color{navyblue}\simeq Cl(0,1).$  Those Clifford algebras displayed above in grey are trivial, and contain no imaginary volume element.  In contrast, each Clifford algebra displayed above in blue is non-trivial, and  has a volume element that squares to -1.

For concreteness, we may take $L_{\mathbb{H}}\simeq Cl(0,2)$ to be generated by $\gamma_1:=L_{\epsilon_1}$ and $\gamma_2:=L_{\epsilon_2},$ leading to an imaginary volume element $\omega:=\gamma_1 \gamma_2 = L_{\epsilon_3}.$
We may take $R_{\mathbb{H}}\simeq Cl(0,2)$ to be generated by $\gamma_1:=R_{\epsilon_1}$ and $\gamma_2:=R_{\epsilon_2},$ leading to an imaginary volume element $\omega:=\gamma_1 \gamma_2 = -R_{\epsilon_3}.$  We may take $B_{\mathbb{H}}\simeq Cl(3,1)$ to be generated by $\gamma_i := L_{\epsilon_1}R_{\epsilon_i},$ for $i\in \{1,2,3\},$ and  $\gamma_4 := L_{\epsilon_2},$ leading to an imaginary volume element $\omega:=\gamma_1 \gamma_2\gamma_3 \gamma_4 = -L_{\epsilon_3}.$   We may take $L_{\mathbb{O}}\simeq Cl(0,6)$ to be generated by $\gamma_j := L_{e_j}$ for $j\in \{1, \dots 6\},$  leading to an imaginary volume element $\omega:=\gamma_1 \gamma_2\gamma_3 \gamma_4\gamma_5 \gamma_6 = L_{e_7}.$  We may take $R_{\mathbb{O}}\simeq Cl(0,6)$ to be generated by $\gamma_j := R_{e_j}$ for $j\in \{1, \dots 6\},$  leading to an imaginary volume element $\omega:=\gamma_1 \gamma_2\gamma_3 \gamma_4\gamma_5 \gamma_6 = -R_{e_7}.$

Careful readers may have noticed that $L_{\C}\simeq R_{\C}\simeq B_{\C}$ are isomorphic to $\textup{End}_{\C}(\C)$, not $\textup{End}_{\R}(\C)$.  It can be confirmed that the results obtained in this article continue to hold when the second line in equations~(\ref{endcliffs}) is replaced by
\begin{equation}\begin{array}{l}\label{endRC}
\textup{End}_{\R}(\C)\simeq M_2(\R) \color{navyblue}\simeq Cl(2,0). \color{black}
\end{array}\end{equation}

\section{Centralizers\label{central}}

Having identified each multiplication algebra with a Clifford algebra, we will now seek out those subalgebras that commute with the volume element $\omega$ (when a nontrivial volume element exists).  To be explicit, we will find the centralizer of the volume element with respect to the full multiplication algebra, $C_{B_{\A}}(\omega)$.   In the case of $Cl(s,t),$ where $s+t$ is even, this amounts to identifying the even part of that Clifford algebra.  Below, we map the original full multiplication algebras to the centralizers of their volume elements.
\begin{equation} \label{centralBox}
\begin{array}{l@{\;}c@{\;}l}
\multicolumn{3}{l}{
\grayunderlinedmath{
{\color{gray} B_{\A}\longmapsto C_{B_{\A}}(\omega)}
}
}
\\[4mm]
B_{\R} 
& \longmapsto 
& \R,
\\[2mm]
B_{\C} 
& \longmapsto 
& \C,
\\[2mm]
B_{\mathbb{H}} 
& \longmapsto 
& M_2(\C),
\\[2mm]
B_{\mathbb{O}} 
& \longmapsto 
& M_4(\C) \longmapsto M_3(\C)\oplus \C.
\end{array}
\end{equation}

The cases of $B_{\R}$ and $B_{\C}$ are trivial, and so the algebra goes unchanged.  The full quaternionic multiplication algebra, $B_{\mathbb{H}},$ reduces to $M_2(\C)$.  The full octonionic multiplication algebra, $B_{\mathbb{O}},$ reduces to $M_4(\C)$ when it is required to commute with either $\omega =L_{e_7}$ or $\omega =-R_{e_7}.$  It reduces further to $M_3(\C)\oplus \C$ when it is required to commute with both.

In short, this establishes a route from 
\begin{equation}\begin{array}{c}\OHCR \vspace{2mm}\\ 
\downarrow \vspace{2mm}\\
\C_{\mathbb{O}} \oplus M_3(\C)_{\mathbb{O}} \oplus M_2(\C)_{\mathbb{H}} \oplus \C_{\C} \oplus \R_{\R}.
\end{array}\end{equation}
\noindent Here, the subscripts refer to the source of the centralizer.  With these endomorphisms reduced in this way, we may now consider $\mathbb{O}$ as a complex vector space $\C_{\mathbb{O}}\oplus \C_{\mathbb{O}}^3,$  $\mathbb{H}$ as a complex vector space $\C_{\mathbb{H}}^2,$ and trivially, $\C$ as a complex vector space, $\C_{\C}.$

Most importantly, this 

\vspace{2mm}

\emph{establishes a possible origin for the diagonal block algebra responsible for the projection lines (particle decomposition) in~\citep{Z5}.}

\vspace{2mm}

That is, readers are encouraged to compare the centralizer algebra
 \begin{equation}\label{DSM}
 \color{navygreen}\Delta_{\textup{SM}}:=\C \oplus M_3(\C) \oplus M_2(\C) \oplus \C \oplus \R \color{black}
 \end{equation}
\noindent to the diagonal blocks of Figure~(\ref{map}).

\section{Electroweak symmetry breaking\label{EWSB}}

Looking back at equations~(\ref{full}), readers may notice a unique property of the full quaternionic multiplication algebra, $B_{\mathbb{H}},$ that is not exhibited in the real, complex, or octonionic sectors.  That is, unlike with these other three sectors, 
\begin{equation} 
B_{\A} \simeq L_{\A} \simeq R_{\A}
\end{equation}
\noindent does not hold in the quaternionic case.  Instead, both $L_{\mathbb{H}}$ and $R_{\mathbb{H}}$ are needed simultaneously in order to describe $\mathbb{H}$'s full multiplication algebra:
\begin{equation} 
B_{\mathbb{H}}  \simeq  L_{\mathbb{H}}\otimes_{\R}R_{\mathbb{H}}.
\end{equation}

Instead of only considering $B_{\mathbb{H}}\simeq Cl(3,1)$ with a single volume element, we may then also consider a second stage whereby $B_{\mathbb{H}}$ is taken as 
\begin{equation} B_{\mathbb{H}}  \simeq  Cl(0,2)\otimes_{\R}Cl(0,2),
\end{equation}
\noindent with one volume element for each factor of $Cl(0,2)$.  Restricting $Cl(0,2)$ to the centralizer of its imaginary volume element leaves us with $\C$ 
\begin{equation} Cl(0,2)  \longmapsto  \C\,
\end{equation}
\noindent so that 
\begin{equation} B_{\mathbb{H}}  \longmapsto  \C\otimes_{\R}\C \simeq \C\oplus \C.
\end{equation}
\noindent Updating equation~(\ref{DSM}) for this second stage gives another centralizer algebra:
\begin{equation}
 \color{navygreen}\Delta_{\textup{LE}}:=\C \oplus M_3(\C) \oplus \C \oplus \C\oplus \C \oplus \R. \color{black}
 \end{equation}
\noindent  It is clear that in going from $\Delta_{\textup{SM}}$ to $\Delta_{\textup{LE}}$ we replace $M_2(\C)$ with $\C\oplus \C.$  That $\Delta_{\textup{SM}}$ and $\Delta_{\textup{LE}}$ lead to the Standard Model's internal symmetries pre- and post-electroweak symmetry breaking is yet to be established in the following sections.

\section{Lie-Jordan splitting\label{split}}

The centralizer algebras $\Delta_{\textup{SM}}$ and $\Delta_{\textup{LE}}$ are given by a series of matrix algebras with natural anti-involution $\dagger,$ and so may be subject to \emph{Lie-Jordan splitting},~\citep{321},\citep{Z5}.  Under $\dagger,$ these $\Delta$ spaces may be split into two parts:
\begin{equation}
\mathcal{L}_{\Delta}:=\{u\in\Delta \hspace{1mm}| \hspace{1mm}u^{\dagger}=-u\}    
 \end{equation}
\noindent and
\begin{equation}
\mathcal{H}_{\Delta}:=\{h\in\Delta \hspace{1mm}|\hspace{1mm} h^{\dagger}=h\}.    
 \end{equation}
For $a, b\in\Delta,$ we define a new product $m(a,b):=ab+ba^{\dagger}.$  Taking $u, u'\in \mathcal{L}_{\Delta}$ and $h,h'\in \mathcal{H}_{\Delta},$ we find:
\begin{equation} \begin{array}{rclll}\label{LJ}
m(h,h')&=&\{\hspace{.5mm}h,\hspace{.5mm}h'\hspace{.5mm}\} &\hspace{1cm} &\in \mathcal{H}_{\Delta},\vspace{2mm}\\
m(u, u')&=&\left[\hspace{.5mm}u,\hspace{.5mm}u'\hspace{.5mm}\right] &\hspace{1cm} &\in \mathcal{L}_{\Delta},\vspace{2mm}\\
m(h, u)&=&\{\hspace{.5mm}h,\hspace{.5mm}u\hspace{.5mm}\} &\hspace{1cm} &\in \mathcal{L}_{\Delta},\vspace{2mm}\\
m(u, h)&=&\left[\hspace{.5mm}u,\hspace{.5mm}h\hspace{.5mm}\right] &\hspace{1cm} &\in \mathcal{H}_{\Delta}.
\end{array}\end{equation}    
\noindent Here, $\{\hspace{.2mm}a,\hspace{.5mm}b\hspace{.5mm}\}:=ab+ba$ defines the Jordan product (up to a factor of 1/2), while $[\hspace{.5mm}a,\hspace{.5mm}b\hspace{.5mm}]:=ab-ba$ defines the Lie product.  The first two lines in equations~(\ref{LJ}) then indicate that the hermitian elements, taken alone, close as a Jordan algebra, and the anti-hermitian elements, taken alone, close as a Lie algebra.  

In a quantum mechanical model, we may interpret $ \mathcal{L}_{\Delta}$ as the system's symmetries, whereas  $\mathcal{H}_{\Delta}$ may be taken to be an algebra of observables.  A set of states may be identified as the set of positive semidefinite elements of $\mathcal{H}_{\Delta}$ with trace equal to 1.  The last line in equation~(\ref{LJ}) then describes how symmetries act on the observables and states.  Readers may then appreciate that \emph{the single action $ab+ba^{\dagger}$ universally encodes how this system's symmetries, observables, and states act on one another.}  See endnote~\citep{Cdensity}.

According to this analysis, we extract the anti-hermitian part of $\Delta_{\textup{SM}}$ as 
\begin{equation}
\color{navygreen}\mathcal{L}_{\Delta_{\textup{SM}}}=\mathfrak{u}(1)_{\mathbb{O}}\oplus\mathfrak{u}(3)_{\mathbb{O}}\oplus \mathfrak{u}(2)_{\mathbb{H}}\oplus \mathfrak{u}(1)_{\C},\color{black}
 \end{equation}
\noindent while we extract the anti-hermitian part of $\Delta_{\textup{LE}}$ as 
\begin{equation}
\color{navygreen}\mathcal{L}_{\Delta_{\textup{LE}}}=\mathfrak{u}(1)_{\mathbb{O}}\oplus\mathfrak{u}(3)_{\mathbb{O}}\oplus \mathfrak{u}(1)_{\mathbb{H}}\oplus \mathfrak{u}(1)_{\mathbb{H}}\oplus \mathfrak{u}(1)_{\C}.\color{black}
 \end{equation}

Readers can confirm that the same result may be found starting with norm-preserving symmetries $\mathfrak{so}(n)$ for $n = dim(\A), $ where $\A = \R, \C, \mathbb{H}, \mathbb{O}.$ 

\section{Extracting symmetries \label{extract}}

It is obvious that $\mathcal{L}_{\Delta_{\textup{SM}}}$ and $\mathcal{L}_{\Delta_{\textup{LE}}}$ contain the Standard Model's pre-Higgs $\mathfrak{g}_{\textup{SM}} := \mathfrak{su}(3) \oplus  \mathfrak{su}(2)\oplus \mathfrak{u}(1),$ and post-Higgs $\mathfrak{g}_{\textup{LE}}:=\mathfrak{su}(3) \oplus \mathfrak{u}(1)$ symmetries,  respectively.  What is not so obvious, however, is how exactly to extract these desired Lie algebras from them.  

The $\OHCR$ structure we have been discussing so far may embed in a variety of models, in a variety of different ways.  In many cases, the needed symmetry constraints might only be realizable once the symmetries are made to materialize as actual particle representations.  We invite readers to explore what is possible.

With this said, in this section, we now isolate $\mathfrak{g}_{\textup{SM}}$ and $\mathfrak{g}_{\textup{LE}}$ according to~\citep{Z5}.

\subsection{$\R,$ $\C,$ $\mathbb{H},$ $\mathbb{O}$ as $\mathbb{Z}_2^n$-graded algebras}

In Sections~\ref{central} and \ref{EWSB}, in order to arrive at $\Delta_{\textup{SM}}$ and $\Delta_{\textup{LE}},$ we found $Cl(s,t)$ structure within certain $\textup{End}_{\R}(\A)$ algebras, whereby the Clifford volume element squares to -1.  We then isolated the subalgebra of $Cl(s,t)$ commuting with that volume element.

The Clifford algebra $Cl(s,t)$ with $s+t=n$ may be viewed as a $\mathbb{Z}_2^n$-graded algebra.  Let us  define a $\mathbb{Z}_2^n$-graded algebra as an algebra $A$ with vector space decomposition 
\begin{equation}
A = \underset{x\in\mathbb{Z}_2^n}\bigoplus A_{x}
\end{equation}
\noindent such that for all $a_x\in A_x$ and $a_y\in A_y,$ multiplication $m$ obeys
\begin{equation}
m(a_x, a_y) \in A_{x+y}.
\end{equation}
\noindent Addition  $x+y$ is performed componentwise, mod 2.  According to this grading, we may now assign to $Cl(s,t)$'s volume element the highest label, namely, $(1, 1, 1, \dots)\in\mathbb{Z}_2^n.$

It was pointed out by Albuquerque and Majid, \citep{shahn}, that $\R,$ $\C,$ $\mathbb{H},$ $\mathbb{O}$ may \emph{also} be viewed as $\mathbb{Z}_2^n$-graded algebras.  Namely, if $\A$ is one of these division algebras of real dimension $2^m,$ then $\A$ is $\mathbb{Z}_2^m$-graded.

\begin{figure}[h]
\begin{center}
\includegraphics[width=9cm]{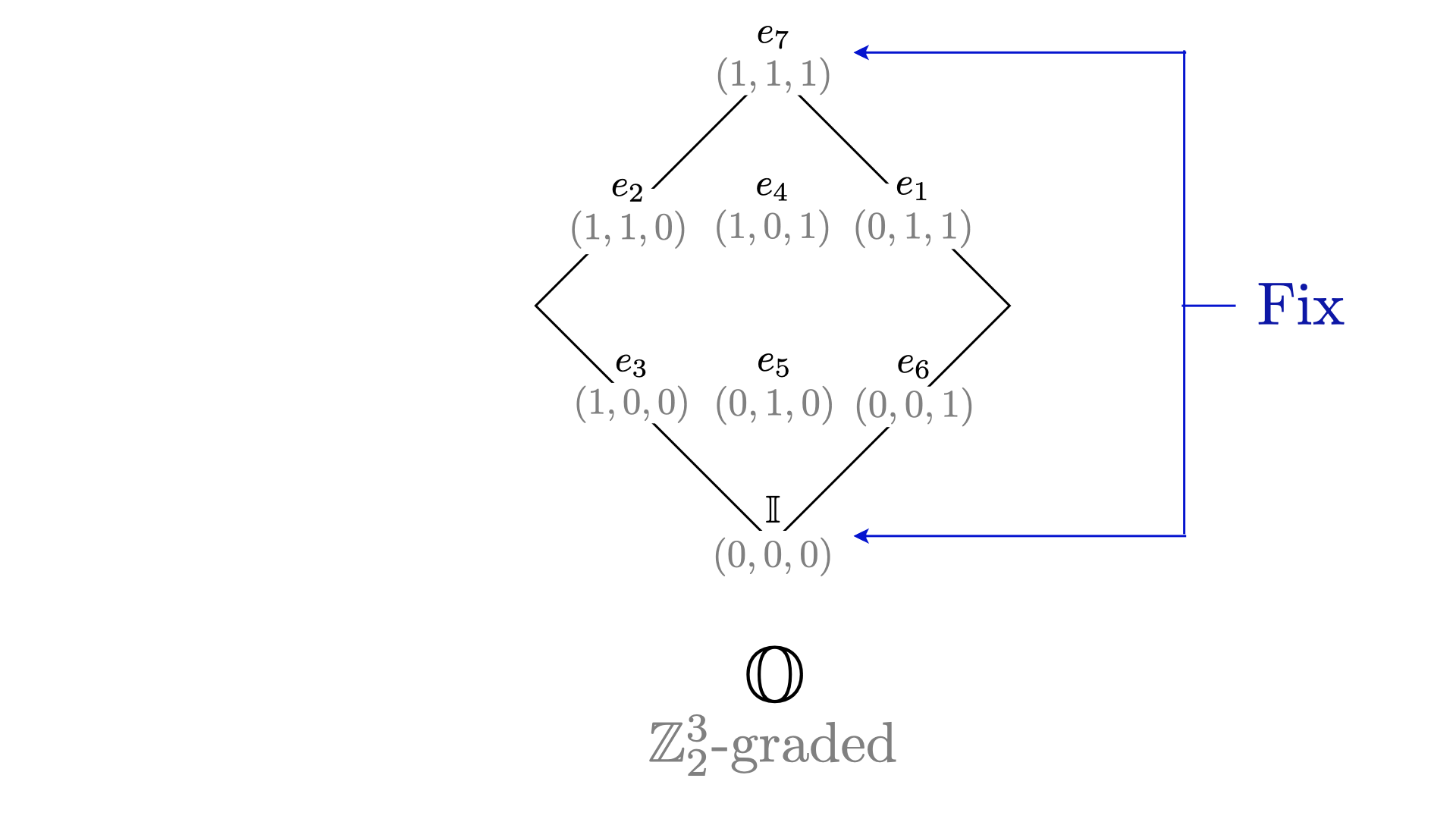}
\caption{\label{Z23}  The octonions constitute a $\mathbb{Z}_2^3$-graded algebra, \citep{shahn}.  Fixing the highest degree element, in this example $e_7,$ helps to isolate the Standard Model's internal symmetries from within $\mathcal{L}_{\Delta_{\textup{SM}}}$ and $\mathcal{L}_{\Delta_{\textup{LE}}}$.}

\includegraphics[width=9.5cm]{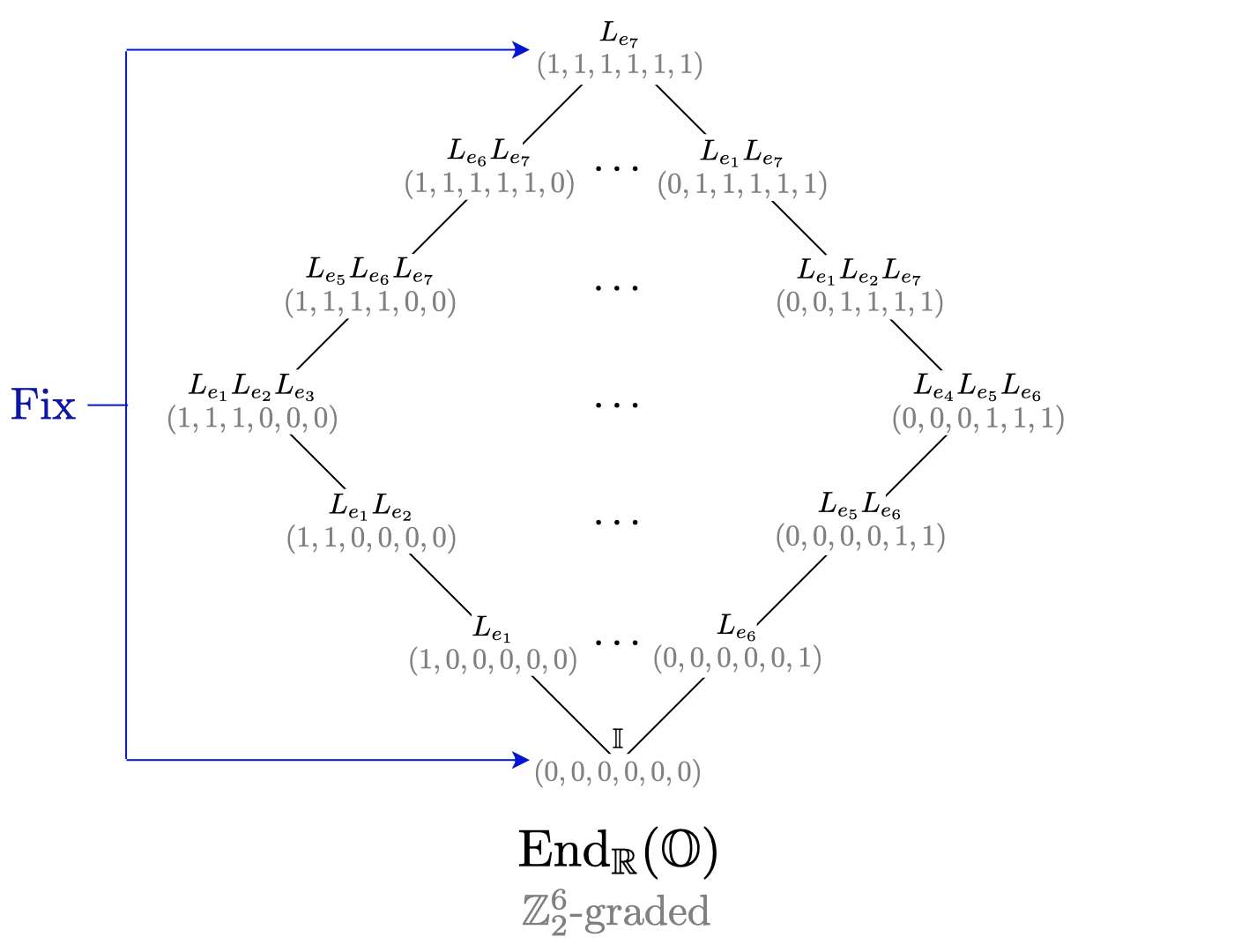}
\caption{\label{Z26}  Similarly, the Clifford algebra $Cl(0,6)\simeq \textup{End}_{\R}(\mathbb{O}) $ constitutes a $\mathbb{Z}_2^6$-graded algebra.  Fixing the highest degree element reduces the real Clifford algebra $Cl(0,6)$ to its even (complex) subalgebra.}
\end{center}\end{figure}

Now, just as we preserved only those symmetries that stabilized the top Clifford volume element, we may again preserve only those symmetries that stabilize this time the top division algebraic element.  Doing so will help us isolate $\mathfrak{g}_{\textup{SM}}$ in $\mathcal{L}_{\Delta_{\textup{SM}}}$ and $\mathfrak{g}_{\textup{LE}}$ in $\mathcal{L}_{\Delta_{\textup{LE}}}.$  See Figures~(\ref{Z23}) and (\ref{Z26}).

\subsection{Two conditions}

Identifying $\mathfrak{g}_{\textup{SM}}$ within $\mathcal{L}_{\Delta_{\textup{SM}}}$ and $\mathfrak{g}_{\textup{LE}}$ within $\mathcal{L}_{\Delta_{\textup{LE}}}$ is now possible in two steps.  At the Lie algebra level, for $\ell\in \mathcal{L}_{\Delta_{\textup{SM}}}$ and  $\ell' \in \mathcal{L}_{\Delta_{\textup{LE}}},$ readers may confirm that 
\begin{equation}\begin{array}{ll}\label{cond}
(I) &\textup{Annihilating highest deg. octonion,}\vspace{2mm}\\
(II) &\textup{Setting}\hspace{2mm}  \textup{Tr} \hspace{.5mm}\!\bigl(J_{\mathbb{O}}\hspace{.5mm}\ell_{\mathbb O}\bigr)
=
\textup{Tr}\hspace{.5mm} \!\bigl(J_{\mathbb{H}}\hspace{.5mm}\ell_{\mathbb H}\bigr)
=
\textup{Tr}\hspace{.5mm} \!\bigl(J_{\mathbb{C}}\hspace{.5mm}\ell_{\mathbb C}\bigr)\vspace{2mm}\\

&\Rightarrow \hspace{5mm}\color{navyblue}\mathcal{L}_{\Delta_{\textup{SM}}} \hspace{1mm}\longmapsto \hspace{1mm}\mathfrak{g}_{\textup{SM}},\color{black}\vspace{6mm}\\

(I) &\textup{Annihilating highest deg. octonion and quaternion,} \vspace{2mm}\\
(II) &\textup{Setting}\hspace{2mm}  \textup{Tr} \hspace{.5mm}\!\bigl(J_{\mathbb{O}}\hspace{.5mm}\ell'_{\mathbb O}\bigr)
=
\textup{Tr}\hspace{.5mm} \!\bigl(J_{\mathbb{H}}\hspace{.5mm}\ell'_{\mathbb H}\bigr)
=
\textup{Tr}\hspace{.5mm} \!\bigl(J_{\mathbb{C}}\hspace{.5mm}\ell'_{\mathbb C}\bigr)\vspace{2mm}\\
&\Rightarrow \hspace{5mm}\color{navyblue}\mathcal{L}_{\Delta_{\textup{LE}}}\hspace{1mm} \longmapsto\hspace{1mm} \mathfrak{g}_{\textup{LE}}. \color{black}
\end{array}\end{equation}
\noindent  Here, $\ell_{\A} := \ell|_{\A}$ and $\ell'_{\A} := \ell'|_{\A}$ are the restrictions of $\ell$ and $\ell'$ to $\A = \C, \mathbb{H}, \mathbb{O},$ and $J_{\mathbb{O}}, J_{\mathbb{H}}, J_{\mathbb{C}}$ are complex structures chosen from the imaginary volume elements of Section~\ref{imagine}.  These trace conditions apply regardless of whether we are taking the real or complex trace.  

Using the conventions of Section~\ref{EWSB}, we now demonstrate for the reader the result of these conditions.

\subsection{Preparing phases} 

The $\mathfrak{su}(3)$ and $\mathfrak{su}(2)$ Lie subalgebras of $\mathcal{L}_{\Delta_{\textup{SM}}}$ satisfy the conditions~(\ref{cond}) automatically.  For what is coming next, it will  be convenient for us to express  $\mathcal{L}_{\Delta_{\textup{SM}}}$'s four $\mathfrak{u}(1)$ symmetries simultaneously, in their hermitian form, as
\begin{equation}\alpha_1 P_{\mathbb{O}_1} + \alpha_2 P_{\mathbb{O}_2} + \beta P_{\mathbb{H}} + \gamma P_{\C}
\end{equation}
\noindent for $  \alpha_1, \alpha_2, \beta, \gamma \in \R.$ The idempotents $P_{\mathbb{O}_1}$ and $P_{\mathbb{O}_2}$ act on the octonions only, while annihilating all else.  According to the conventions of Section~\ref{imagine}, they are given by 
\begin{equation}P_{\mathbb{O}_1}:=\frac{1}{2}\left(\hspace{.5mm}\mathbb{I}_{\mathbb{O}}-L_{e_7}R_{e_7}\right), \hspace{6mm}P_{\mathbb{O}_2}:=\frac{1}{2}\left(\hspace{.5mm}\mathbb{I}_{\mathbb{O}}+L_{e_7}R_{e_7}\right)  .
\end{equation}
\noindent The idempotent $P_{\mathbb{H}}$ acts as the identity on $\mathbb{H}$, while annihilating all else.  Similarly, $P_{\mathbb{C}}$ acts as the identity on $\C$ while annihilating all else.

It may likewise be confirmed that the $\mathfrak{su}(3)$  Lie subalgebra of $\mathcal{L}_{\Delta_{\textup{LE}}}$ satisfies the conditions~(\ref{cond}) automatically.  For what is coming next, it will be convenient for us to express $\mathcal{L}_{\Delta_{\textup{LE}}}$'s five $\mathfrak{u}(1)$ symmetries simultaneously, in their hermitian form, as 
\begin{equation}  \alpha_1' P_{\mathbb{O}_1} + \alpha_2' P_{\mathbb{O}_2} + \beta_1' P_{\mathbb{H}_1} + \beta_2' P_{\mathbb{H}_2}+ \gamma' P_{\C}    
\end{equation}
\noindent for  $\alpha_1',\alpha_2', \beta_1',\beta_2', \gamma'\in \R.$  The idempotents $P_{\mathbb{H}_1}$ and $P_{\mathbb{H}_2}$ act on the quaternions only, while annihilating all else.  According to the conventions of Section~\ref{imagine}, they are given by 
\begin{equation}P_{\mathbb{H}_1}:=\frac{1}{2}\left(\hspace{.5mm}\mathbb{I}_{\mathbb{H}}-L_{\epsilon_3}R_{\epsilon_3}\right), \hspace{6mm}P_{\mathbb{H}_2}:=\frac{1}{2}\left(\hspace{.5mm}\mathbb{I}_{\mathbb{H}}+L_{\epsilon_3}R_{\epsilon_3}\right)  .
\end{equation}
\noindent From here, we may now finally solve for $Y$ and $Q.$

\subsection{$\mathfrak{g}_{\textup{SM}}$}

Consider $\ell\in\mathcal{L}_{\Delta_{\textup{SM}}} = \mathfrak{u}(1)_{\mathbb{O}}\oplus\mathfrak{u}(3)_{\mathbb{O}}\oplus \mathfrak{u}(2)_{\mathbb{H}}\oplus \mathfrak{u}(1)_{\C}.$  At the Lie algebra level, condition (I): fixing the highest degree element in $\mathbb{O},$ becomes a condition that $\ell$ annihilate that element.   Concretely,
\begin{equation}
(I) \hspace{5mm} \ell (e_7) = 0 \hspace{2mm}\Rightarrow \hspace{2mm}\alpha_1 = 0,
\end{equation}
\noindent so that now $\ell\in \mathfrak{u}(3)_{\mathbb{O}}\oplus \mathfrak{u}(2)_{\mathbb{H}}\oplus \mathfrak{u}(1)_{\C}.$  At the Lie algebra level, condition (II) is a condition of traces.  Concretely,
\begin{equation}\begin{array}{l}
(II) \hspace{5mm}\textup{Tr} \hspace{.5mm}\!\bigl(J_{\mathbb{O}}\hspace{.5mm}\ell_{\mathbb O}\bigr)
=
\textup{Tr}\hspace{.5mm} \!\bigl(J_{\mathbb{H}}\hspace{.5mm}\ell_{\mathbb H}\bigr)
=
\textup{Tr}\hspace{.5mm} \!\bigl(J_{\mathbb{C}}\hspace{.5mm}\ell_{\mathbb C}\bigr) \vspace{2mm}\\
\hspace{2cm}\Rightarrow \hspace{2mm}3\alpha_2 = 2\beta = \gamma,
\end{array}\end{equation}
\noindent so that now $\ell\in \mathfrak{su}(3)_{\mathbb{O}}\oplus \mathfrak{su}(2)_{\mathbb{H}}\oplus \mathfrak{u}(1)_{Y},$ with weak hypercharge Y surviving as
 \begin{equation} \color{navygreen}Y = \frac{1}{3} P_{\mathbb{O}_2} + \frac{1}{2}P_{\mathbb{H}} + P_{\C}.\color{black}
 \end{equation}

\subsection{$\mathfrak{g}_{\textup{LE}}$}

From what we found in Section~\ref{EWSB} on electroweak symmetry breaking, readers may foresee that it is possible to find $Q$ via the same procedure as $Y$.  The only difference is in the first step where we isolate the highest degree element in both $\mathbb{O}$ \emph{and} $\mathbb{H}$.  

Consider $\ell'\in\mathcal{L}_{\Delta_{\textup{LE}}} = \mathfrak{u}(1)_{\mathbb{O}}\oplus\mathfrak{u}(3)_{\mathbb{O}}\oplus \mathfrak{u}(1)_{\mathbb{H}_1}\oplus\mathfrak{u}(1)_{\mathbb{H}_2}\oplus \mathfrak{u}(1)_{\C}.$  At the Lie algebra level, condition (I): fixing the highest degree element in $\mathbb{O}$ and $\mathbb{H},$ becomes a condition that $\ell'$ annihilate those elements.   Concretely,
\begin{equation}
(I) \hspace{5mm} \ell' (e_7) = \ell' (\epsilon_3)= 0  \hspace{2mm}\Rightarrow \hspace{2mm}\alpha_1' =  \beta_1' = 0,
\end{equation}
\noindent so that now $\ell'\in \mathfrak{u}(3)_{\mathbb{O}}\oplus \mathfrak{u}(1)_{\mathbb{H}_2}\oplus \mathfrak{u}(1)_{\C}.$  At the Lie algebra level, condition (II) is a condition of traces.  Concretely,
\begin{equation}\begin{array}{l}
(II) \hspace{5mm}\textup{Tr} \hspace{.5mm}\!\bigl(J_{\mathbb{O}}\hspace{.5mm}\ell'_{\mathbb O}\bigr)
=
\textup{Tr}\hspace{.5mm} \!\bigl(J_{\mathbb{H}}\hspace{.5mm}\ell'_{\mathbb H}\bigr)
=
\textup{Tr}\hspace{.5mm} \!\bigl(J_{\mathbb{C}}\hspace{.5mm}\ell'_{\mathbb C}\bigr) \vspace{2mm}\\
\hspace{2cm}\Rightarrow \hspace{2mm}3\alpha_2' = \beta_2' = \gamma',
\end{array}\end{equation}
\noindent so that now $\ell'\in \mathfrak{su}(3)_{\mathbb{O}}\oplus \mathfrak{u}(1)_{Q},$ with electric charge Q surviving as
 \begin{equation}  \color{navygreen}Q = \frac{1}{3} P_{\mathbb{O}_2} + P_{\mathbb{H}_2} + P_{\C}.\color{black}
 \end{equation}

\subsection{Some observations\label{obs}}

\bf Observation (a): \rm Throughout this work, one finds that several complex structures are often simultaneously available for use.  For example, one may choose either $L_{e_7}$ or $R_{e_7}$ in order to define $\mathfrak{su}(3)$ triplet structure of quarks.  For the same $\mathfrak{su}(3)$ generators acting on the same vector space, whether the vector space is interpreted as a quark or anti-quark \emph{depends on the choice of complex structure.}  That is, $L_{e_7}$ vs $R_{e_7}.$

This might have two important consequences.  First of all, for those familiar, it may enable helicity to be described more easily in algebraic models, \citep{Z5}.  But more emphatically: we propose investigating \emph{whether or not choice of complex structure could ultimately resolve the matter anti-matter asymmetry problem (baryon asymmetry problem).}  

\bf Observation (b):  \rm It is possible to confirm that the $Y$ and $Q$ operators found in this article are equivalent to those found in~\citep{Z5}.  However, the form of those charge operators in~\citep{Z5} was more complicated.  The equivalence can be understood by noticing that an overall phase generator (proportional to the identity) cancels when the operators are applied to the particle blocks.   That is, given a commutator action, only the difference between the left- and right-actions matters.  One achievement of this present construction is then the presentation of these $Y$ and $Q$ operators in a particularly simple form.

\bf Observation (c):  \rm Given that $Y$ and $Q$ are hermitian operators, they then reside within $\mathcal{H}_{\Delta_{\textup{SM}}}$ and $\mathcal{H}_{\Delta_{\textup{LE}}}$ of Section~\ref{split}.  With the \emph{standard normalization} of weak hypercharge and electric charge, these operators each, surprisingly, exhibit the form
\begin{equation} \sum \frac{1}{n} \mathbb{I}_{n\times n},
\end{equation}
\noindent where $n$ refers to complex dimension.  Curiously, each term then represents the density matrix corresponding to a maximally mixed state.  That is, a state with maximal von Neumann entropy, and the least prior knowledge.

\subsection{Traceless $Y_0$ and $Q_0$}

In Observation (b), it was noted that an overall phase does not affect the particle spectrum in Figure~(\ref{map}).  Thus, we are free to find traceless versions of $Y$ and $Q.$  When $\mathbb{V}$ is 16$\hspace{.5mm}\R$ or 8$\hspace{.5mm}\C$ dimensional, as we will see shortly, it so happens that these traceless versions, $Y_0$ and $Q_0$ take the form
\begin{equation}\begin{array}{lllll}
Y_0 &=& Y -\frac{3}{8}\hspace{.5mm}\mathbb{I}, \vspace{2mm}\\
 Q_0 &=& Q -\frac{3}{8}\hspace{.5mm}\mathbb{I},
\end{array}\end{equation}
\noindent where
\begin{equation} 
\frac{\textup{Tr}(Y)}{\textup{dim}(\mathbb{V})} = \frac{\textup{Tr}(Q)}{\textup{dim}(\mathbb{V})}= \frac{3}{8}.
\end{equation}
\noindent The result holds regardless of whether we are taking the real or complex trace, $\textup{Tr}_{\R}$ or $\textup{Tr}_{\C}.$

\section{Nested Cayley-Dickson models}

We have now established a route from $\OHCR$ to $\mathfrak{g}_{\textup{SM}}= \mathfrak{su}(3)\oplus \mathfrak{su}(2)\oplus\mathfrak{u}(1)$ and $\mathfrak{g}_{\textup{LE}}= \mathfrak{su}(3)\oplus\mathfrak{u}(1).$   From here, it is upon us to explain how the structure leads to \emph{Nested Cayley-Dickson models}.

\subsection{Popular embedding spaces}

It is indeed possible to construct a model based on $\OHCR$ and its vector-space endomorphisms alone: $\textup{End}_{\R}(\OHCR)\simeq M_{15}(\R)$.  However, we point out that this $\OHCR$ vector space decomposition appears naturally in algebras that many researchers have already been studying.

$\OHCR$ is 15$\hspace{.5mm}\R$ dimensional, and may embed in several well-studied 16$\hspace{.5mm}\R$ dimensional spaces, which we refer to generically as $\mathbb{V}$.  Take, for example, $\mathbb{O}\oplus\mathbb{O},$ \citep{Silagadze}, \citep{bryant}, the sedenions, $\mathbb{S},$ \citep{NAAP}, \citep{Jens_sed}, \citep{Carlos2019sed}, \citep{masi},  or $\CO \simeq \mathbb{O}\oplus i\mathbb{O}$ thought of as a real vector space, \citep{Gen}, \citep{321}, \citep{gsm}, \citep{singh_v}, \citep{boyle1}.  

The $\OHCR$ structure may also be relevant for models based on the 32$\hspace{.5mm}\R$ dimensional complex sedenions, $\C\otimes\mathbb{S},$ \citep{Weng}, \citep{niels_sedenion}.  However, as mentioned earlier, we are interested in augmenting   $\OHCR$ by only one real dimension, since $\textup{End}_{\R}(\mathbb{V})\simeq Cl(0,8)$ may then connect to the ideas of Bott Periodic particle physics, \citep{Bott1}, \citep{Bott2}, \citep{Z5}.

\subsection{Nested Cayley-Dickson embeddings}

Consider some 16$\hspace{.5mm}\R$ dimensional vector space $\mathbb{V}$ given by $\mathbb{O}\oplus\mathbb{O},$ the sedenions, $\mathbb{S},$ $\CO \simeq \mathbb{O}\oplus i\mathbb{O},$ or some other suitable construct.  In such suitable cases, we may decompose $\mathbb{V}$ as a vector space in the following way: 
\begin{equation}\label{Vsimp}
\mathbb{V} \hspace{.5mm}= \hspace{.5mm}e_i \mathbb{O} \hspace{.5mm}\oplus \hspace{.5mm}e_5 \mathbb{H} \hspace{.5mm}\oplus \hspace{.5mm}e_6 \mathbb{C} \hspace{.5mm}\oplus\hspace{.5mm} e_7 \mathbb{R} \hspace{.5mm}\oplus\hspace{.5mm} \R,
\end{equation}
\noindent where $e_i$ is the identity in the case of $\mathbb{O}\oplus\mathbb{O},$ $e_i$ is the sedenionic imaginary element $e_8$ in the case of $\mathbb{S},$ and $e_i$ is $i$ in the case of $\CO \simeq i\mathbb{O}\oplus \mathbb{O}.$  For clarity, we may write this decomposition as 
\begin{equation}\label{num_decomp}
\mathbb{V} \hspace{.5mm}= \hspace{.5mm}e_i \mathbb{O}_{12\dots 7} \hspace{.5mm}\oplus\hspace{.5mm} e_5 \mathbb{H}_{267} \hspace{.5mm}\oplus \hspace{.5mm}e_6 \mathbb{C}_7\hspace{.5mm} \oplus \hspace{.5mm}e_7 \mathbb{R} \hspace{.5mm}\oplus \hspace{.5mm}\R,
\end{equation}

\noindent so as to specify that $\mathbb{C}$ in equation~(\ref{Vsimp}) represents elements $r_0+r_7e_7$ for $r_0, r_7\in\R,$ $\mathbb{H}$ represents elements $r_0'+r_j'e_j$ for $j\in\{2,6,7\},$ and $r_0', r_j'\in\R,$   $\mathbb{O}$ represents elements $r_0''+r_k''e_k$ for $k\in\{1,2,\dots 7\}$, and $r_0'', r_k''\in\R.$

\begin{figure}[h]
\begin{center}
\includegraphics[width=9cm]{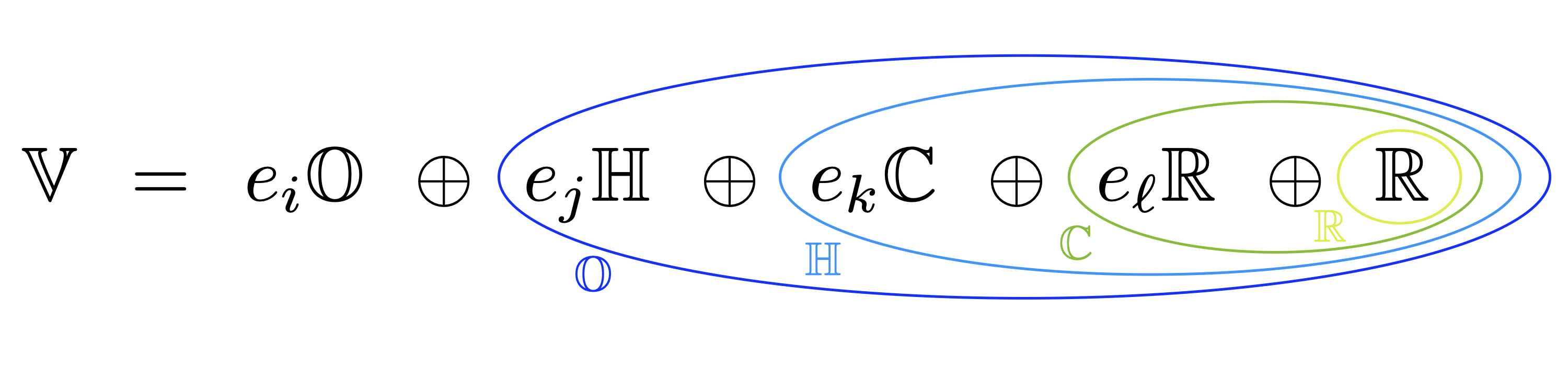}
\caption{\label{comet}  Nested division algebraic inclusions $\mathbb{V} \supset \mathbb{O} \supset  \mathbb{H} \supset  \mathbb{C}\supset \mathbb{R}.$  When $\mathbb{V} = \mathbb{S}$ the full sequence forms a Cayley-Dickson tower.  Note that $i, j, k, \ell$ must take on specific values (as in equation~\ref{Vsimp}) in order to support nested Cayley-Dickson algebras. }
\end{center}\end{figure}

From here, it is straightforward to see how $\mathbb{V}$ admits a series of nested inclusions:
\begin{equation}
\mathbb{V} \supset \mathbb{O} \supset  \mathbb{H} \supset  \mathbb{C}\supset \mathbb{R}.
\end{equation}
\noindent In the case of $\mathbb{V} = \mathbb{S},$ the full series forms a nested Cayley-Dickson tower.  See Figure~(\ref{comet}).

We then define the $\R$-linear embedding $\iota_{\mathbb{V}}:\mathbb{O}_{12\dots7}\oplus \mathbb{H}_{267}\oplus \C_7 \oplus \R \longrightarrow \mathbb{V}$ by 
\begin{equation}\label{iota}
\iota_{\mathbb{V}}(o,h,c,r) = e_i o + e_5 h + e_6 c + e_7 r,
\end{equation}
\noindent $\forall(o,h,c,r)\in\mathbb{O}_{12\dots7}\oplus \mathbb{H}_{267}\oplus \C_7 \oplus \R.$  It is easy to confirm from equations~(\ref{num_decomp}) and (\ref{iota}) that $\iota_{\mathbb{V}}$ is a vector space embedding that maps a 15$\hspace{.5mm}\R$ dimensional space (injectively) into a 16$\hspace{.5mm}\R$ dimensional space $\mathbb{V} = \textup{im}(\iota_{\mathbb{V}})\oplus \R$.  There is no claim that $\iota_{\mathbb{V}}$ is an algebra homomorphism, and it need not be for our purposes.

Given the vector space decomposition $\mathbb{V} = \textup{im}(\iota_{\mathbb{V}})\oplus \R,$ it is straightforward to identify an induced action of $\mathfrak{g}_{\textup{SM}}$ on $\mathbb{V}.$  Let $\iota_{\mathbb{V}}^{-1}$ denote the inverse of $\iota_{\mathbb{V}},$ mapping $\textup{im}(\iota_{\mathbb{V}}) \rightarrow \OHCR.$  Then $X\in \mathfrak{g}_{\textup{SM}}$ is represented by $\rho_{\mathbb{V}}(X)$ on $\mathbb{V}$ as 
\begin{equation} \begin{array}{l}
\rho_{\mathbb{V}}(X)|_{\textup{im}(\iota_{\mathbb{V}})} =\iota_{\mathbb{V}}\hspace{1mm} \rho_{0}(X) \hspace{1mm}\iota_{\mathbb{V}}^{-1},
\end{array}\end{equation}
\noindent while being represented trivially on $\R \subset \mathbb{V}$ as $\rho_{\mathbb{V}}(X)|_{\R} = 0.$ \emph{Here, $\rho_0$ is the original representation of $\mathfrak{g}_{\textup{SM}}$ on $\OHCR$ that we established in Section~\ref{extract}.}  

With this definition, 
\begin{equation}
\iota_{\mathbb{V}} \circ \rho_0(X) = \rho_{\mathbb{V}}(X) \circ \iota_{\mathbb{V}}
\end{equation}
\noindent $\forall X \in \mathfrak{g}_{\textup{SM}},$  so that $\iota_{\mathbb{V}}$ is a $\R$-linear $\mathfrak{g}_{\textup{SM}}$-equivariant embedding.  The same results hold for $\mathfrak{g}_{\textup{LE}}$.  

With $\rho_{\mathbb{V}}$ defined on all of $\mathbb{V},$ it is easy to identify a global complex structure, e.g. $L_{e_7},$ on $\mathbb{V}$ under which $\rho_{\mathbb{V}}$ is complex-linear.

Intuitively, readers may see that $\iota_{\mathbb{V}}^{-1}$ acts to remove the Cayley-Dickson imaginary units $e_i, e_j, e_k, e_{\ell}$ from $\mathbb{V}$ in Figures~\ref{comet}-\ref{endoC}.  $\rho_0$ then acts on the bare division algebras $\mathbb{O}, \mathbb{H}, \mathbb{C}, \R,$  and $\iota_{\mathbb{V}}$ replaces the Cayley-Dickson imaginaries again.  This goes to say that the $\mathfrak{g}_{\textup{SM}}$ representation one finds within the $\C\oplus M_3(\C) \oplus M_2(\C) \oplus \C \oplus \C$ diagonal blocks of $\textup{End}_{\C}(\mathbb{V})$ may be interpreted as ultimately acting directly on $\mathbb{O}, \mathbb{H}, \mathbb{C}, \R$ behind the Cayley-Dickson imaginary units $e_i, e_j, e_k, e_{\ell}$.

\subsection{Endomorphic models}

\begin{figure}[h]
\begin{center}
\includegraphics[width=9.5cm]{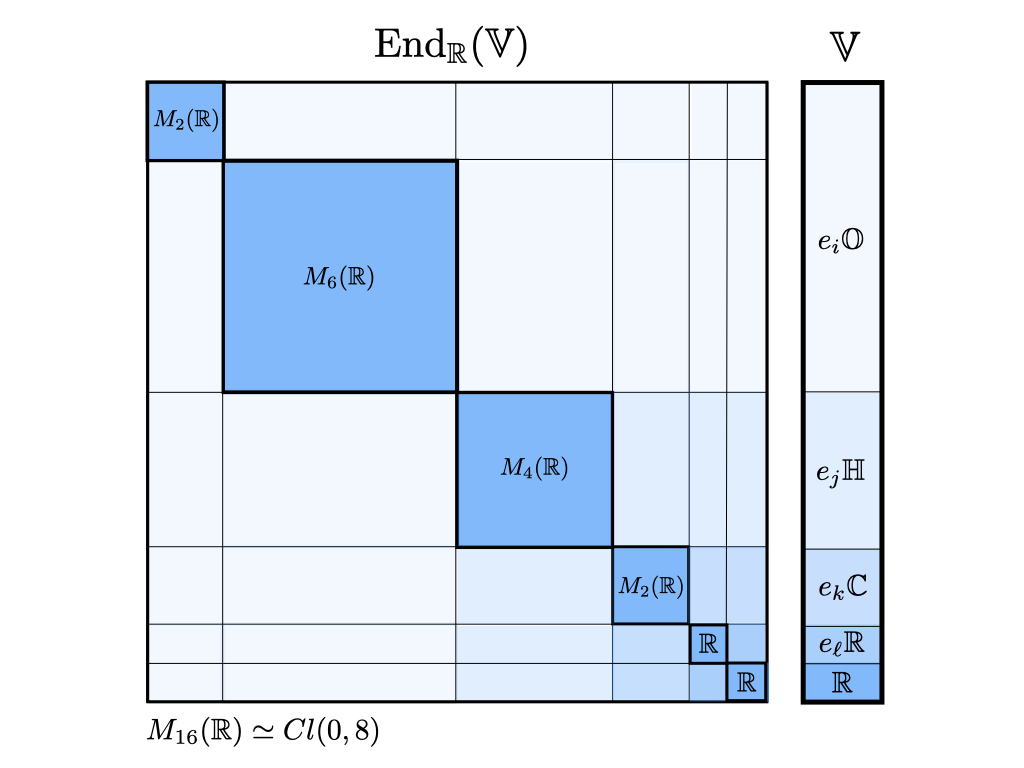}
\caption{\label{endoR}Recasting~\citep{Z5} in terms of $\textup{End}_{\R}(\mathbb{V})\simeq Cl(0,8)$  may link that work to the ideas of Bott Periodic Particle Physics, \citep{Bott1}, \citep{Bott2}, \citep{Z5}.  The particle block decomposition found there may be seen to result in part from the $\R\subset\C\subset\mathbb{H}\subset\mathbb{O}\subset\mathbb{V}$ nested Cayley-Dickson substructure of $\mathbb{V}.$}
\end{center}\end{figure}

\begin{figure}[h]
\begin{center}
\includegraphics[width=9.5cm]{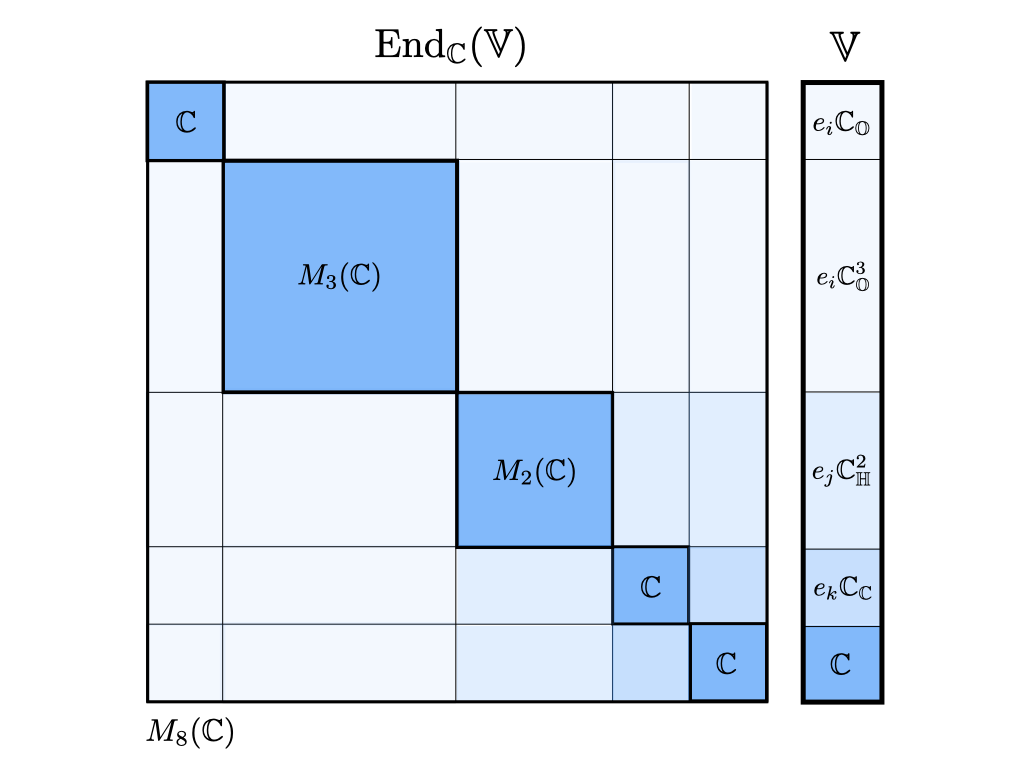}
\caption{\label{endoC}  Thought of as an 8$\hspace{.5mm}\C$ vector space, $\mathbb{V}$ plays the role of a left $\textup{End}_{\C}(\mathbb{V})$ module.  Readers may confirm that the Peirce decomposition induced by the nested Cayley-Dickson embeddings $\R\subset\C\subset\mathbb{H}\subset\mathbb{O}\subset\mathbb{V}$ enables a decomposition into Standard Model irreps analogous to those already established in~\citep{Z5}.  }
\end{center}\end{figure}

As mentioned earlier, 16$\hspace{.5mm}\R$ dimensional algebras are of special interest, as they may serve as a left $Cl(0,8)$-module in a model of \emph{Bott Periodic Particle Physics}, \citep{Bott1}, \citep{Bott2}, \citep{Z5}.  In future work, we consider $Cl(0,8)\simeq \textup{End}_{\R}(\mathbb{V})\simeq M_{16}(\R) $ and its reduction to  $\textup{End}_{\C}(\mathbb{V})\simeq M_{8}(\C) $ once a complex structure is implemented.  Please see Figures~(\ref{endoR}) and (\ref{endoC}).  Special care will be taken with regard to the nuances of multiplication algebras, as explained in this text.

At this stage, readers may  confirm that the Peirce decomposition of $\textup{End}_{\C}(\mathbb{V})$ in Figure~\ref{endoC} enables a decomposition into particle irreps that matches what is seen in Figure~\ref{map} and already established in reference \citep{Z5}.  The main difference is that in this case, the counting is on-shell, and $\textup{End}_{\C}(\mathbb{V})$ is not hermitian, but rather isomorphic to $M_8(\C).$

It is clear that the vector space decomposition of $\mathbb{V}$ in Figure~\ref{comet} induces the Peirce decompositions seen in Figures~\ref{endoR} and \ref{endoC}.  These Peirce decompositions dictate much of the model's structure by restricting the allowed symmetries, and defining particle blocks.  It is straightforward to confirm that these Peirce decompositions persist under multiplication of elements in $\textup{End}(\mathbb{V}).$   In fact, they turn these endomorphism algebras into $\mathbb{Z}_2^5$-graded algebras, as was the case in~\citep{Z5}.  See endnote~\citep{Z4}.  Put more bluntly, even though each component in $\mathbb{V}$'s decomposition does not close as a subalgebra, the endomorphisms that comprise our model do indeed support a particle decomposition that persists under multiplication and addition.

\subsection{ $\mathbb{R}\oplus\mathbb{C}\oplus\mathbb{H}\oplus\mathbb{O}$ inside $\RCHO$}

With regard to choosing a specific algebra for $\mathbb{V},$ naively, the sedenions $\mathbb{S},$ provide the most obvious candidate.  This option then embodies one current direction of inquiry.

With this said, $\CO,$ as a part of the larger $\RCHO$ algebra is also worth exploring.  Taken as a real vector space, $\CO$ can serve as a left $Cl(0,8)$-module.  As mentioned earlier in this text, the remaining endomorphisms of $\mathbb{H},$ $\textup{End}_{\R}(\mathbb{H}),$ may be taken to be $Cl(3,1).$  It would then be straightforward to set up a model based on $\textup{End}_{\R}(\RCHO) \simeq Cl(0,8) \otimes_{\R} Cl(3,1)$ with $Cl(0,8)$ representing internal degrees of freedom, and $Cl(3,1)$ representing spacetime degrees of freedom.  Reducing $Cl(3,1)$ to $M_2(\C)\simeq \CLtwo$ via a complex structure would then give us access to both $Cl(0,8)$ and  $\CLtwo:$ the two repeating elements of real and complex Bott Periodicity.  This $Cl(0,8) \otimes_{\R} Cl(s_0,t_0)$ decomposition was already proposed in \citep{Bott1}, \citep{Bott2}, \citep{Z5}. 

More directly, we may consider $\RCHO$  as a module of $\textup{End}_{\C}(\RCHO) \simeq Cl(0,8) \otimes_{\R} \CLtwo\simeq Cl(1,10) \simeq \CLten$, as per the proposals  \citep{321}, \citep{Bott2}, and \citep{Z5}.   Ultimately, we seek out a particle physics realization of the Tenfold Way, \citep{AZ10}, \citep{Kit10}, \citep{RSFL10}, \citep{Bott2}, \citep{Z5}.

\subsection{Connection to exceptional Lie algebras \label{e8}}

The $\OHCR$ vector space decomposition may play a role in models based on exceptional Lie algebras.   Consider the $\mathfrak{e}_8$ vector space decomposition
\begin{equation}
\mathfrak{e}_8 \hspace{.5mm}= \hspace{.5mm}\mathfrak{tri}(\mathbb{O}_1) \hspace{.5mm}\oplus \hspace{.5mm}\mathfrak{tri}(\mathbb{O}_2) \hspace{.5mm}\oplus \hspace{.5mm}3\cdot \mathbb{O}_1\otimes  \mathbb{O}_2,
\end{equation}
\noindent where $\mathbb{O}_1$ and $\mathbb{O}_2$ may represent octonions or split octonions.  From this framework of nested Cayley-Dickson algebras, we propose decomposing one copy, $\mathbb{O}_2,$ in terms of its nested Cayley-Dickson substructure.  For example, in the case that $\mathbb{O}_2 = \mathbb{O},$
\begin{equation}
\mathbb{O}_1\otimes  \mathbb{O}_2 \hspace{2mm} \mapsto \hspace{2mm} \mathbb{O}_1\otimes \left( e_j\mathbb{H}\oplus e_k \C \oplus e_{\ell}\R \oplus \R  \right).
\end{equation}
\noindent This suggests a possible way to transplant the $\OHCR$-based $\mathfrak{g}_{\textup{SM}}$-extracting procedure into an $\mathfrak{e}_8$ framework.  For some suitably chosen weak hypercharge, $Y$, one would expect two of the $\mathbb{O}_1\otimes  \mathbb{O}_2$ particle representations to match those of the outer-off-diagonal blocks in Figure~(\ref{map}).  Some significant overlap should also occur between $\mathfrak{tri}(\mathbb{O}_1) \hspace{.5mm}\oplus \hspace{.5mm}\mathfrak{tri}(\mathbb{O}_2)$ and the diagonal blocks of Figure~(\ref{map}).

\subsection{The Division Algebraic Hopf Tree}

We close by pointing out an appearance in this model of the fully connected division algebraic tree of Hopf fibrations involving the four parallelizable spheres $S^7, S^3, S^1, S^0$.

\begin{figure}[h]
\begin{center}
\includegraphics[width=7.5cm]{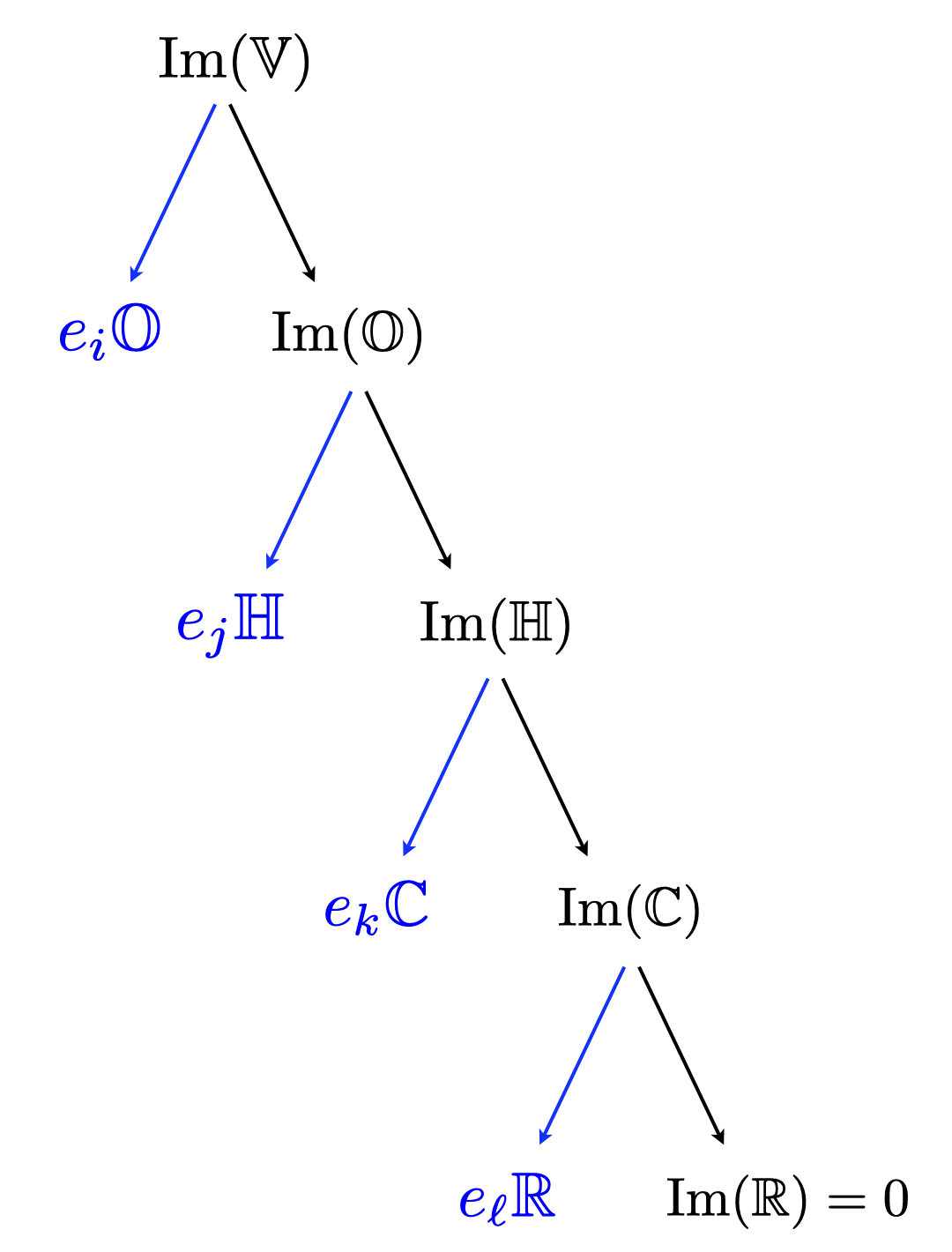}
\caption{\label{CDtree}The Peirce decomposition central to this $Cl(0,8)$ model is largely induced by the decompositions within the \emph{Cayley-Dickson tree} shown above.  }
\end{center}\end{figure}

The Peirce decomposition central to this project may be viewed as being induced by the vector space decomposition of `Im($\mathbb{V}$)' as shown in Figure~\ref{CDtree}.  For efficiency we write `Im($\mathbb{V}$)' as shorthand for the vector space $\textup{Im}(\mathbb{V}):=e_i\mathbb{O} \oplus \textup{Im}(\mathbb{O})$.  Figure~\ref{CDtree} depicts this decomposition in what we will refer to as an (imaginary) \emph{Cayley-Dickson tree}.

\begin{figure}[h!]
\begin{center}
\includegraphics[width=7.5cm]{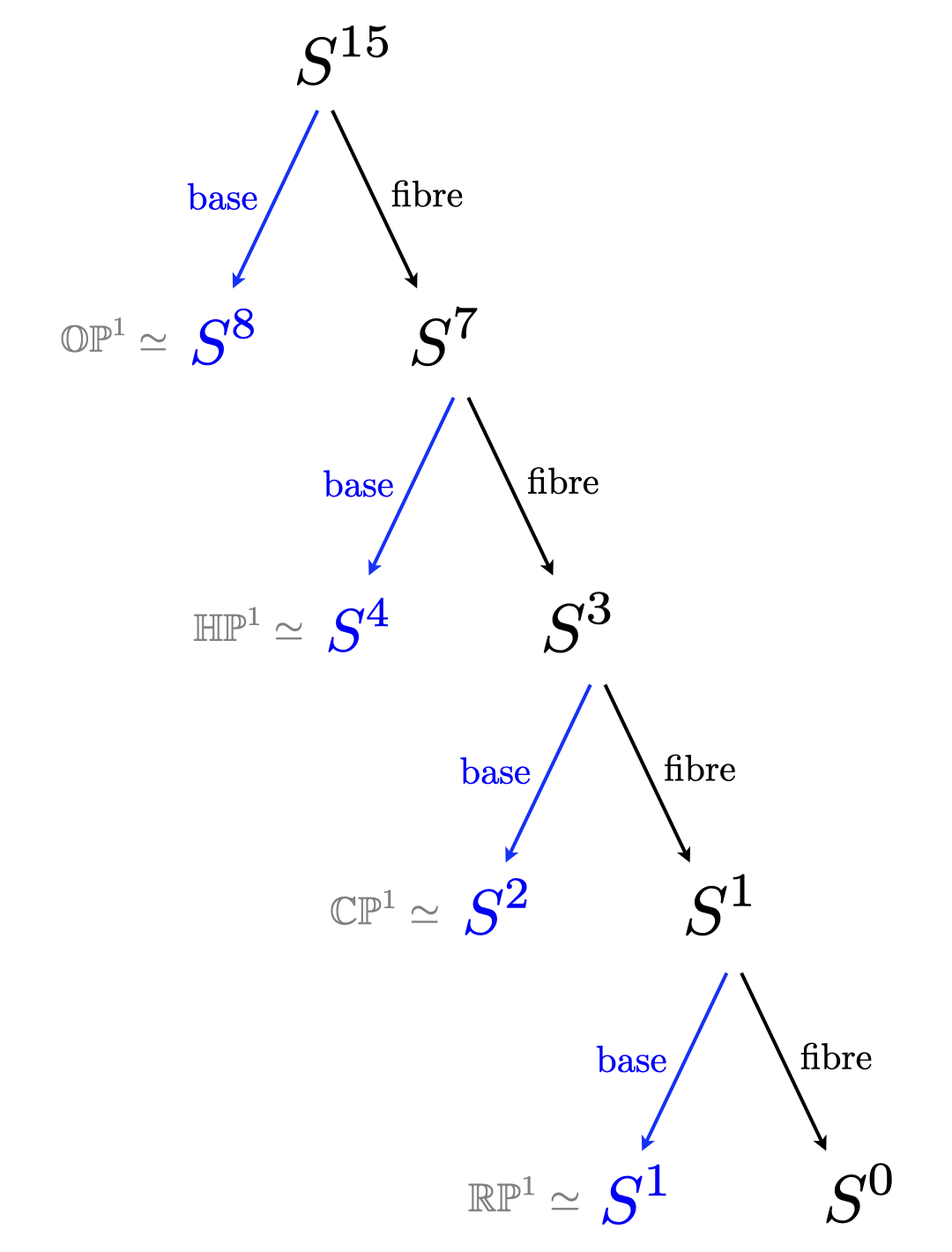}
\caption{\label{Hopftree}The fully connected tree of Hopf fibrations involving the four parallelizable spheres $S^7,S^3, S^1, S^0$.}
\end{center}\end{figure}

Noting the dimension counts at each level of Figure~\ref{CDtree}, and also the identities 
\begin{equation}\begin{array}{lll}
S(\mathbb{V}) \simeq S^{15}\vspace{2mm}\\
S(\mathbb{O}) \simeq S^{7}\vspace{2mm}\\
S(\mathbb{H}) \simeq S^{3}\vspace{2mm}\\
S(\mathbb{C}) \simeq S^{1}\vspace{2mm}\\
S(\mathbb{R}) \simeq S^{0},
\end{array}\end{equation}
\noindent one may infer the existence of a connection between the Cayley-Dickson tree of Figure~\ref{CDtree} and the division algebraic Hopf fibration tree of Figure~\ref{Hopftree}.  Here $S(\mathbb{W}):= \{ w\in \mathbb{W} \hspace{1mm}: \hspace{1mm}|w| = 1 \}.$


The connection between Figure~\ref{CDtree} and Figure~\ref{Hopftree} may indeed be made concrete. Let $\A = \mathbb{O}, \mathbb{H}, \C, \R $ and  $\A' = \A \oplus e_m\A$.  For $a_1+e_ma_2\in\A',$ with $|a_1|^2+|a_2|^2=1,$ consider the Hopf map $h_{\A}:S(\A') \rightarrow \A \mathbb{P}^1$ given by
\begin{equation}
h_{\A}(a_1+e_ma_2) =  \left(\hspace{.5mm}|a_1|^2- |a_2|^2, \hspace{.5mm}2a_1\bar{a}_2\hspace{.5mm}\right),
\end{equation}
\noindent where $\bar{a}_2$ is the usual division algebraic conjugate of $a_2$.  

From here it can be confirmed that the tangent spaces of these base spheres at their respective north poles $p_{\A}:=h_{\A}(1)$ may be described by the $e_m\A$ vector spaces written in blue in the Cayley-Dickson tree of Figure~\ref{CDtree}.
\begin{equation}\begin{array}{ccccccccc}

 T_1S^{15} &=&&&&&&&

\vspace{2mm}\\
 
 \textup{Im}(\mathbb{V}) &=&e_i \mathbb{O} &\oplus&e_j \mathbb{H} &\oplus&e_k \mathbb{C} &\oplus&e_{\ell} \mathbb{R}

\vspace{2mm}\\

&\simeq& T_{p_{\mathbb{O}}}\hspace{.5mm}S^8 &\oplus&T_{p_{\mathbb{H}}}\hspace{.5mm}S^4 &\oplus&T_{p_{\mathbb{C}}}\hspace{.5mm}S^2 &\oplus&T_{p_{\mathbb{R}}}\hspace{.5mm}S^1

\vspace{2mm}\\

  &\simeq&T_{p_{\mathbb{O}}}\hspace{.5mm}\mathbb{O}\mathbb{P}^1 &\oplus&T_{p_{\mathbb{H}}}\hspace{.5mm}\mathbb{H}\mathbb{P}^1 &\oplus&T_{p_{\mathbb{C}}}\hspace{.5mm}\mathbb{C}\mathbb{P}^1 &\oplus &T_{p_{\mathbb{R}}}\hspace{.5mm}\mathbb{R}\mathbb{P}^1.

 \vspace{2mm}\\

\end{array}\end{equation}
\noindent  Please see endnote~\citep{OO}.  

This leads us to the interpretation that the internal gauge bosonic degrees of freedom in our model are vector space endomorphisms mapping these tangent spaces to themselves.  The internal fermionic and Higgs degrees of freedom are vector space homomorphisms mapping one tangent space to another.

The idea that parallelizable spheres, division algebraic projective spaces, and division algebraic Hopf fibrations underlie an internal geometry for the Standard Model is not new.  Consider for example works by Duff, Koh, Nilsson, Pope, \citep{Duff1}, \citep{Duff2}, Grossman, Kephart,  Stasheff, \citep{GKS}, Woit, \citep{Woit}, Waldron and Joshi, \citep{WJ},  Davis, \citep{Davis1}, \citep{Davis2},  G. Avila, S. J. Castillo, J. A. Nieto, \citep{Mex}, Wolk, \citep{wolk2020}, \citep{wolk2023}, Szangolies, \citep{sz}.   

We show for the first time that a majority of Standard Model internal irreps coincide with maps between north-pole tangent spaces of the four Hopf base spheres: $S^1,$ $S^2,$ $S^4,$ $S^8$.

\section{Outlook}

In this article, we have demonstrated that much of the Standard Model's internal structure may be seen to arise from:  

$$\begin{array}{l}
(1) \textup{ The vector space endomorphisms of } \OHCR, \vspace{1mm}\\

(2)  \textup{ The nested embeddings of } \R\subset \C \subset \mathbb{H} \subset \mathbb{O} \subset \mathbb{V}, \vspace{1mm}\\

(3)   \textup{ North-pole tangent spaces of the Hopf base spheres: }  \vspace{1mm}\\
\hspace{.55cm}S^1, S^2, S^4, S^8. \vspace{1mm}
\end{array}$$

 In upcoming work, we demonstrate how this model's Peirce decomposition (particle irreps) arises in part from the sequence of nested Clifford algebras of the Bott Periodic clock.  That is, $\{Cl(0,n)\}$ or $\{Cl(n,0)\}$ for $n=1, 2, \dots 8.$  This finding further develops the research efforts of \emph{Bott Periodic Particle Physics}, \citep{Bott1}, \citep{Bott2}, \citep{Z5}.

\begin{acknowledgments}  

\noindent \emph{Elin, Janina, Yasmin, my bright, my bright, my bright.} \rm

This material was presented for the Exceptional Structures and Standard Model Workshop, University of Edinburgh, 2026.07.13.

The author is grateful for feedback from Latham Boyle, Dariusz Chruscinski, Mia Hughes, Rob Klabbers, Jens K\"{o}plinger, Kaushal Kumar, Beth Romano, Gniewko Sarbicki, Shadi Tahvildar-Zadeh, Graham White, Jorge Zanelli.  A sincere thanks goes to J. Nielsen for turning my attention to the geometry specific to the $\R\subset \C \subset \mathbb{H} \subset \mathbb{O} \subset \mathbb{V}$ model described in this article.  Most of all, my gratitude to David Chester and Mia Hughes for your friendship and courage.

This work was graciously supported by the VW Stiftung Freigeist Fellowship, and Humboldt-Universit\"{a}t zu Berlin.


\end{acknowledgments}

\medskip


\begin{thebibliography}{11}
\bibitem{endom} As we will see, a base algebra's multiplication algebra does not always amount to the entire set of vector space endomorphisms.  This is true in particular when one considers real vector space endomorphisms on a complex base algebra.
\bibitem{Cdensity}  Particle irreps were identifiable as hermitian objects in~\citep{Z5}.  In this current article, these irreps live in the full complex part.
\bibitem{Z4}  It is in fact possible to describe these $\mathbb{Z}_2^5$-graded algebras slightly more efficiently as $\mathbb{Z}_2^4$-graded algebras.
\bibitem{OO}  In the case of $\mathbb{V} = \mathbb{O} \oplus \mathbb{O},$ we take $1\in \mathbb{V}$ to mean 1 in the first copy of $\mathbb{O}$ and 0 in the second copy of $\mathbb{O}$.
\end{thebibliography}
\end{document}